\documentstyle[aps,pre,epsf ,multicol]{revtex}

\begin{document}
\title{\bf Persistence Exponents and the Statistics of Crossings and 
Occupation Times for Gaussian Stationary Processes}
\author{G. M. C. A. Ehrhardt$^1$, Satya N. Majumdar$^2$, 
and Alan J. Bray$^1$} 
\address{$^1$Department of Physics and Astronomy, University of 
Manchester, Manchester M13 9PL \\
$^2$Laboratoire de Physique Quantique (UMR C5626 du CNRS), Universit'e 
Paul Sabatier, 31062 Toulouse Cedex, France}

\maketitle

\begin{abstract}
\end{abstract}

\begin{abstract}
\noindent We consider  the persistence probability, the occupation-time
distribution and the distribution of  the number of zero crossings for
discrete  or  (equivalently)  discretely sampled  Gaussian  Stationary
Processes   (GSPs)   of   zero    mean.    We   first   consider   the
Ornstein-Uhlenbeck  process,  finding  expressions  for the  mean  and
variance  of  the  number  of  crossings and  the  `partial  survival'
probability.  We then elaborate  on the correlator expansion developed
in an earlier paper (G. C. M. A. Ehrhardt and A. J. Bray, Phys.\ Rev.\
Lett.\ 88, 070602  (2001)) to calculate  discretely sampled persistence
exponents of GSPs  of known correlator by means  of a series expansion
in the correlator.  We apply this method to the processes 
$d^n x/dt^n=\eta(t)$
with $n  \ge 3$, incorporating an  extrapolation of the  series to the
limit of continuous sampling.  We then extend the correlator method to
calculate  the occupation-time  and crossing-number  distributions, as
well  as  their  partial-survival  distributions and  the  means  and
variances of  the occupation  time and number  of crossings.  We apply
these general methods to  the $d^n x/dt^n=\eta(t)$ processes for $n=1$
(random  walk), $n=2$  (random acceleration)  and larger  $n$,  and to
simple  diffusion from  random initial  conditions in  1-3 dimensions.
The results  for discrete sampling  are extrapolated to  the continuum
limit where possible.
\end{abstract}

\section{Introduction} \label{Introduction}

Stochastic processes driven by Gaussian  white noise have a wide range
of  applications in  the physical  sciences and  beyond,  ranging from
Brownian  motion to  options  pricing.   Here we  focus  on two  basic
properties  of  a  stochastic  Gaussian  time series:  the  number  of
crossings of  the mean value of  the series, and the  fraction of time
for which the series is above  its mean value.  The former, termed the
crossing  number,   has  long  been  of  interest   to  engineers  and
mathematicians\cite{rice,bendat,blakeandlindsey}, and more recently to
physicists\cite{mandbpartialsurvival,majumdarpsieqphiiplusphiiminus1}.
The  latter, termed  the occupation  time,  has also  been studied  by
mathematicians  for  a   long  time\cite{levy,kac,lamperti}  for  both
Gaussian and non-Gaussian stochastic processes and has recently seen a
revival  in the  physics community  in the  context  of nonequilibrium
systems\cite{dornic,newman,dharandmajumdardxresidenz,various,newmanandloinaz}.
The  occupation-time distribution  for  a stochastic  process is  also
important due to  its potential applications in a  variety of physical
systems  which include  optical imaging\cite{weiss},  analysis  of the
morphology  of  growing  surfaces\cite{tnd}, analysis  of  temperature
fluctuations in  weather records\cite{mandbresidenzetc}, in disordered
systems\cite{mc} and also due to the connection between the occupation
time in certain discrete sequences and spin-glass models\cite{md}.

Of particular  interest is  a limiting case  of these  two properties,
namely   the   probability,   termed   the   persistence   probability
\cite{review}, that the time series  is always above its mean value up
to  time  $T$.   The  latter,  for stationary  Gaussian  time  series,
typically decays as $\exp(-\theta T)$ for $T$ large, with the exponent
$\theta$  in  general  taking  a  nontrivial  value.   For  continuous
processes  $X(T)$,   the  Independent  Interval   Approximation  (IIA)
\cite{IIA1,IIA2}   may  be  used   to  calculate   approximately  (for
continuous sampling) the asymptotic (i.e.\ large-$T$) forms of some of
the probability  distributions above.  This approach, which  makes the
(generally  invalid)  assumption   that  the  time  intervals  between
zero-crossings are statistically independent, is surprisingly accurate
in   many  cases.    However,   the  IIA   involves  an   uncontrolled
approximation,  which cannot be  improved upon  in general,  and whose
numerical accuracy is  hard to estimate.  Until now,  the IIA has been
the  only general analytical  technique available.  In the  absence of
exact general results, calculations of probability distributions exist
only for  certain specific processes, although  a short-time expansion
for a general process has been developed \cite{Zeitak}.

Both  continuous   and  intrinsically  discrete   time-series  can  be
studied.  In  this paper  we  consider  discrete-time  sampling of  an
underlying continuous Gaussian  stationary process (GSP), $X(T)$, with
zero  mean,  unit  variance,  and  known correlator  $C(T)  =  \langle
X(T)X(0) \rangle$. We sample this process every time step $\Delta\!T$,
and  study  the discrete-time  series  $X(i\Delta\!T)$.   This is,  of
course, completely equivalent to  studying a discrete process with the
same correlators $C(j \Delta\!T)$.   In \cite{mbe,ebm} we have studied
the persistence of a discretely  sampled random walk and of a randomly
accelerated particle, both of which can be mapped to a GSP by a change
of variables.  In \cite{ctl} we calculated, using  a series  expansion 
in the correlator  $C(j\Delta\!T)$, the persistence probability for an
arbitrary  discretely-sampled GSP.  By extrapolating  to the  limit in
which the  time between samplings  tends to zero, we  obtained results
for  several  continuum  processes.  Thus the  results  developed  for
discretely sampled  processes may, for  sufficiently smooth processes,
be extended to give results for continuous-time processes. 

Before going further, we illustrate the main ideas by considering the 
simple example of a stochastic process, the continuous-time random 
walk described by the Langevin equation
\begin{equation}  
\dot x=\xi(t)
\label{dummy1}  
\end{equation} 
where $\xi(t)$ is Gaussian white noise, $\left< \xi(t)
\right>=0$, and $\left< \xi(t) \xi(t')\right>=2 D \delta(t-t')$.
The probability that $x>0$ up to time $t$ decays as
$t^{-\theta}$ where the `persistence exponent' is $\theta=1/2$.
This process is not stationary since its correlator, 
\begin{equation}  
C(t_1,t_2)= \left< (x(t_1)-\left<x\right>) 
(x(t_2)-\left<x\right>) \right> = 2D\,{\rm min}(t_1,t_2)
\label{dummy2}  
\end{equation} 
does not depend only on the time difference, $|t_1 -t_2|$.  Note that,
since  the process  is Gaussian,  it  is completely  specified by  its
correlator and  mean.  We can map  the random walk onto a stationary
process  by a  change of  variables;  we change  to logarithmic  time,
$T=\ln t$, and to a normalized process $X(T)$ via
\begin{equation}  
X(T) \equiv { x(t) - \left< x(t) \right> \over {\sqrt {\left< x(t)^2
\right> -\left< x(t) \right>^2} }}\ ,
\label{dummy3}  
\end{equation}
obtaining the equation 
\begin{equation}
{{d X}\over {dT}}=-{1 \over 2} X + \eta(T),
\label{dummy4}
\end{equation}
where $\eta(T)$ is again a Gaussian white noise, and $X(T)$
has zero mean.  The process $X(T)$ is the Ornstein-Uhlenbeck process. 
It is stationary, with correlator 
\begin{equation}  
C(T_1,T_2)= \left< X(T_1) X(T_2) \right> = \exp(-|T_1-T_2|/2)\ .
\label{dummy5}  
\end{equation} 
An  equivalent mapping  can be  made for  any  non-stationary Gaussian
processes  for  which  the  correlator  has  the  form  $C(t_1,t_2)  =
t_1^\alpha g(t_1/t_2)$. Thus although  here we only attempt to analyze
stationary processes,  the results  are more widely  applicable.  Note
that the  exponential decay, $\exp(-\theta  T)$, is equivalent  to the
power-law  decay,  $t^{-\theta}$,  hence the  terminology `persistence
exponent' for $\theta$.

The occupation time is the number $s$  of positive (say)
values obtained  from the $n$ measurements of  $X(T)$. Let $R_{n,s}$ be 
the probability distribution of $s$ for given $n$ and $r = s/n$ 
be the {\em fraction} of measurements that are positive. In the limit 
$n \to \infty$, $s \to \infty$, with $r=s/n$ fixed, $R_{n,s}$ has the 
asymptotic form $R_{n,s} \sim [\rho(r)]^n = \exp[-\theta_D(r) T]$, 
where $T = n\Delta\!T$ and $\theta_D(r) = -\ln[\rho(r)]/\Delta\!T$. 
Here $\theta_D(1)=\theta_D(0)$ (all or none of the measurements positive)  
is the usual discrete persistence exponent introduced in \cite{mbe}. 
In a similar way we can define $P_{n,m}$ to be the probability of 
observing $m$ zero-crossings in $n$ measurements. If now $r=m/n$ and we 
take the limit $n \to \infty$, $m \to \infty$, holding $r=$ fixed, we 
find $P_{n,m} \sim [\rho(r)]^n = \exp[-\theta_D(r)T]$. Here $\theta_D(0)$ 
(no crossings) corresponds to the usual discrete persistence. Although we  
use the same symbols $\rho(r)$ and $\theta_D(r)$ for the occupation-time 
and crossing problems, it should be clear from the context which problem 
we are referring to. 

In this paper  we extend the technique of  \cite{ctl} to calculate the
exponents $\theta_D(r)$, or equivalently the functions $\rho(r)$,  for 
the occupation-time distribution and the  distribution of  crossings for
arbitrary   discrete   or   discretely-sampled   Gaussian   stationary
processes.  The technique gives the exponents as a series expansion in
the  correlators   $C(j  \Delta\!T)$  up  to   $C(10  \Delta\!T)$  and
$C(\Delta\!T)^{10}$,  i.e.\ 10th  order.  For  the calculation  of the
persistence exponent we work to 14th order.  The results work well for
$C(j  \Delta\!T)$  small,  i.e.\  the  time  between  samplings  large
compared  to the  correlation  time of  the  stationary process.   For
certain processes we  are able to extrapolate the  series to the limit
$\Delta\!T \to  0$, thus obtaining  values of the  continuum exponents
that compare favourably with those  predicted by the IIA when measured
against exact or numerical results.

The  layout   of  this   paper  is  as   follows:  In   Part  \ref{The
Ornstein-Uhlenbeck Process} we consider the Ornstein-Uhlenbeck process
introduced  above,  this  being  perhaps  the simplest  of  GSPs.   By
extending the `matrix method' developed in \cite{mbe} we find the mean
and variance  of the distribution of zero-crossings  as a perturbation
expansion to  high accuracy.  We also  use another method  to find the
same results.  The case of an unstable potential (obtained by changing
the sign of the drift term) is also considered.  We extend the concept
of partial survival  \cite{mandbpartialsurvival} to discrete sampling,
calculating $F_n(p)$, the probability of surviving to $n$ samplings if
each detected zero-crossing is survived with probability $p$.

In  Part  \ref{The  Correlator  Expansion}  we apply  and  extend  the
correlator expansion method  of calculating the persistence exponents,
occupation-time distribution and crossing distribution of an arbitrary
discretely-sampled  GSP.  The general  method is then applied  to some
specific  examples of interest,  and the  results extrapolated  to the
limit of continuum sampling where possible.

In  section  \ref{An  introduction  to the  correlator  expansion}  we
introduce the  correlator expansion first  developed to 14th  order in
\cite{ctl} as  a method of  calculating discretely-sampled persistence
exponents.   We  explain  this  technique  more  fully  including  the
extrapolation  to   the  continuum  limit   using  constrained  Pad\'e
approximants, which allows rather accurate calculation of the standard
persistence exponents.   In \cite{ctl}  this technique was  applied to
the case of the random acceleration process and also to diffusion from
random initial conditions in 1-3 dimensions.  Here we also apply it to
the  class  of  processes  $d^n  x/dt^n=\eta(t)$  where  $\eta(t)$  is
Gaussian  white noise.   The $n=1,2$  cases  are the  random walk  and
random acceleration  problems already  studied. Here we  calculate the
persistence exponents  for larger values  of $n$ and  show numerically
that  $\theta_n -  \theta_{\infty} \propto  1/n$ for  $n$  large.  The
results are  compared to the  predictions of the  Independent Interval
Approximation (IIA).

In Sections \ref{Residenz-time distribution} and \ref{Residenz Partial
Survival}  we  extend  the   correlator  expansion  to  calculate  the
occupation-time distribution, $R_{n,s}$, this being the probability of
$s$  positive measurements  in $n$  samplings,  to 10th  order in  the
correlator.   This  gives,   in  particular,   the  variance   of  the
occupation-time   distribution   which   we   calculate  in   a   more
straightforward  way  as  a   check.   We  define  a  partial-survival
occupation probability,  $P_n(p)$, as the probability  of surviving to
$n$ samplings  if each positive sampling is  survived with probability
$p$.   This is  also the  generating function  for $R_{n,s}$.  We find
$P_n(p)$ to 14th order in the  correlator. We apply the results to the
following  five GSPs:  the  random  walk, where  the  results of  Part
\ref{The  Ornstein-Uhlenbeck  Process} are  used  as  a  check of  the
method,  the random  acceleration  problem and  diffusion from  random
initial conditions  in 1, 2  and 3 dimensions.  Extrapolations  to the
continuum are included.

In Sections  \ref{Distribution of Crossings}, \ref{The Variance of the
Number  of  Crossings}  and  \ref{Distribution  of  Crossings  Partial
Survival}  we further  extend  the correlator  expansion to  calculate
$P_{n,m}$, the probability of $m$ detected crossings in $n$ samplings.
This is found  to 10th order in the correlator and  also enables us to
calculate  the partial-survival probability  and the  moments  of the
crossing  distribution.  We  apply  the  result to  the  five GSPs  of
section  \ref{Residenz-time  distribution},   and  also  to  the  $d^n
x/dt^n=\eta(t)$  processes for  $n>2$  and to  an intrinsically
discrete   process   for   which   the   exact   results   are   known
\cite{majumdarpsieqphiiplusphiiminus1}.     Extrapolations    to   the
continuum are included  and we compare continuum results  with the IIA
and also, for the random acceleration partial-survival problem and the
intrinsically discrete  process, to  the exact solutions.   In Section
\ref{The Variance of  the Number of Crossings} we  show the result for
the mean number of detected crossings.  We also derive the variance as
a series expansion in the  correlator, the coefficients of which agree
with those of  the previous section. We conclude  with brief summary of
the results.

\part{The Ornstein-Uhlenbeck Process} \label{The Ornstein-Uhlenbeck Process}

In  Part  \ref{The  Ornstein-Uhlenbeck  Process}  we  will  study  the
detected crossings of the Ornstein-Uhlenbeck process.  Beside being of
interest in  its own right, this  will illustrate some  of the methods
used later and also provide several checks on the correlator expansion
of Part \ref{The Correlator Expansion}.  Furthermore, here we are able
to calculate probability distributions  starting at a certain position
$X_0$,   rather  than   just   the  long-time   or  stationary   state
distributions.

Consider the stationary Gaussian Markov process evolving via
the Langevin equation,
\begin{equation}
{{d X}\over {dT}}=-\mu X + \eta(T),
\label{lange}
\end{equation}
where $\eta(T)$ is a white noise with mean zero and 
correlator, $\langle \eta(T) \eta(T')\rangle =
2D\delta(T-T')$.  This is the Ornstein-Uhlenbeck process
whose persistence exponent for discrete sampling was
calculated in \cite{mbe}.

Integrating  equation (\ref{lange}), we get
\begin{equation}
X(T)= X_0 e^{-\mu T} +e^{-\mu T}\int_0^T \eta(T_1)e^{\mu T_1}dT_1,
\label{xt}
\end{equation}
where $X_0 = X(T=0)$. Let $T=n\Delta T$. Then the mean
position $\langle X_n\rangle $ after $n$ steps starting
initially at $X_0$ is given from equation (\ref{xt}) by,
\begin{equation}
\langle X_n\rangle = X_0 e^{-\mu n \Delta\!T} = X_0 a^n,
\label{avx}
\end{equation}
where $a=e^{-\mu \Delta\!T}$. Similarly one finds that the
correlation function is given by,
\begin{equation}
\langle [X_n-\langle X_n\rangle][X_{m}-\langle X_{m}\rangle]\rangle 
= D' ( a^{|n-m|} -a^{n+m}),
\label{nnc}
\end{equation}
where $D'=D/\mu$.  Thus in the stationary state, $n \to
\infty$, $m \to \infty$ with $n-m$ fixed, this process has
mean zero and a correlator $C(T_1,T_2) \equiv \,\left
< X(T_1) X(T_2) \right > \,$ given by
\begin{equation}
C(T_1,T_2) =  C(|T_2-T_1|)=D' e^{-\mu |T_2-T_1|}.
\label{10}
\end{equation}

\section{Discrete Backward Fokker-Planck Equation} 
\label{Discrete Backward Fokker Planck Equation} 

Let $Q_n(m,X)$ be the probability that starting at $X$ at
$T=0$, the process, {\em when sampled only at the discrete
points}, changes sign $m$ times within $n$ discrete steps.
Note that the probability $P_{n,m}$ of observing $m$ sign
changes in $n$ measurements (as defined in the introduction)
can be simply obtained from $Q_n(m,X)$ via the relation,
$P_{n,m}= \int_{-infty}^{\infty}Q_n(m,X)p_0(X)dX$ where
$p_0(X)$ is the initial distribution of $X$ which we will
take as the stationary distribution of $X$.
One can write down a recursion relation for $Q_n(m,X)$, valid 
for $X>0$, by noting that at the first step either the process 
changes sign or it does not. 
\begin{equation}
Q_{n+1}(m,X)= \int_0^{\infty} Q_n(m,Y) G(Y,\Delta\!T|X,0)
dY+\int_{-\infty}^{0} Q_n(m-1,Y) G(Y,\Delta\!T|X,0) dY,
\label{dummy6}
\end{equation}
where $G(Y,\Delta\!T|X,0)$ is the probability of going from
$X$ to $Y$ in a time $\Delta\!T$, given by
\begin{equation}
G(Y,\Delta\!T|X,0) = \frac{1}{\sqrt{2\pi D'
(1-a^2)}}\,e^{-\frac{(Y-a X)^2}{2D'(1-a^2)}}.
\label{propagator}
\end{equation}
Using the rescaled variables $x=X/{\sqrt {D'(1-a^2)}}$ and
$y=Y/{\sqrt {D'(1-a^2)}}$, and making use of the symmetry
$Q_n(m,-x)=Q_n(m,x)$ we get,
\begin{equation}
Q_{n+1}(m,x)= {1\over {\sqrt {2\pi}}}\int_0^{\infty} \left[
Q_n(m,y)e^{-(y-ax)^2/2} +
Q_n(m-1,y)e^{-(y+ax)^2/2}\right]dy\ .
\label{recursion}
\end{equation}
Note that for $m=0$ this reduces to the persistence problem
studied in \cite{mbe}, whilst for $m=n$ we have the
`alternating persistence' problem \cite{mbe}. 
      
We define the generating function, $F_n(p,x)=\sum_{m=0}^n
Q_n(m,x)p^m$.  The generating function has a physical
interpretation: If one considers that with every detected
change of sign a particle survives with probability $p$
(partial survival), then $F_n(p,x)$ is precisely the
survival probability of the particle. Note that for $p=1$ 
this probability is $1$ whilst for $p=0$ we recover the usual 
discrete persistence.

Multiplying equation (\ref{recursion}) by $p^m$ and summing
over $m$ we get,
\begin{equation}
F_{n+1}(p,x)= {1\over {\sqrt {2\pi}}}\int_0^{\infty}F_n(p,y)
\left[ e^{-(y-ax)^2/2} + p e^{-(y+ax)^2/2}\right]dy.
\label{rec2}
\end{equation} 

Let $E_n(x)=\sum_{m=0}^{\infty}
mQ_n(m,x)=dF_n(p,x)/dp|_{p=1}$ denote the expected number of
sign changes up to $n$ steps starting at $x$ at $T=0$.
Taking the derivative with respect to $p$ and putting $p=1$
in equation (\ref{rec2}), it follows that $E_n(x)$ satisfies
the recursion relation,
\begin{equation}
E_{n+1}(x)={1\over     {\sqrt    {2\pi}}}\int_0^{\infty}E_n(y)
\left[ e^{-(y-ax)^2/2} +  e^{-(y+ax)^2/2}\right]dy +
{1\over {2}} {\rm erfc}\left ( { {ax}\over {\sqrt 2}}\right),
\label{enx}
\end{equation}
where ${\rm erfc}(x)$ is the standard complimentary error
function and the recursion in equation (\ref{enx}) starts
with the initial condition, $E_0(x)=0$.  Also note that
$G_n(x)=\sum_{m=0}^{\infty}
m(m-1)Q_n(m,x)=d^2F_n(p,x)/dp^2|_{p=1}$ satisfies the
recursion,
\begin{equation}
G_{n+1}(x)={1\over {\sqrt {2\pi}}}\int_0^{\infty}G_n(y)
\left[ e^{-(y-ax)^2/2} + e^{-(y+ax)^2/2}\right]dy + {2\over
{\sqrt {2\pi}}}\int_0^{\infty} dy E_n(y) e^{-(y+ax)^2/2},
\label{gnx}
\end{equation}
with the initial condition $G_0(x)=0$, and where $E_n(x)$ is
given by the solution of equation (\ref{enx}). In order to
calculate the average number of crossings and the variance
around this average, we need to solve the two integral
equations (\ref{enx}) and (\ref{gnx}).

If we can obtain the solution of $E_n(x)$ from equation
(\ref{enx}), then we need to average over the distribution
of the initial position $x$ to obtain ${\langle m\rangle}_n
=\int_{-\infty}^{\infty} E_n(x)p_0(X)dX$ where $p_0(X)$ is
the initial distribution of the position $X$ and $x=X/{\sqrt
{D'(1-a^2)}}$.  For $a<1$, i.e. $\mu>0$, if we choose
$p_0(X)$ to be the stationary distribution of the process,
$p_0(X)={1\over {\sqrt {2\pi D'}}}e^{-X^2/2D'}$, which
corresponds to waiting an infinitely large number of steps 
before starting the measurements, then ${\langle m\rangle}_n$ 
can be computed exactly from equation (\ref{enx}).  Multiplying
equation (\ref{enx}) by this $p_0(X)$ and integrating over
$X$ we get

\begin{equation}
{\langle m\rangle}_{n+1}={\langle m\rangle}_n + \lambda,
\label{avgm}
\end{equation}
where $\lambda= {1\over
{2}}-{1\over{\pi}}{\sin}^{-1}(a)$. Using ${\langle
m\rangle}_0=0$, we get exactly,

\begin{equation}
{\langle m\rangle}_{n} = n \lambda
\label{avgm1}
\end{equation}
which is the random walk case of the general result ${\langle
m\rangle}_{n}/n = 1/2 -(1/\pi) \sin^{-1} C(\Delta\!T)$
\cite{barbe}.  Note that for $\Delta\!T \to 0$ this reduces
to Rice's formula, while for $\Delta\!T \to \infty$ it
reduces to $n/2$ as expected since the values of $X$ at the
discrete points become statistically independent.  Note that
for $a>1$, there is no stationary distribution as this
corresponds to an unstable potential.

Note however that to compute $g_n={\langle m(m-1)\rangle}_n$ 
by a similar method, we need to know the full function $E_n(x)$, 
that is we need to solve the full integral equation (\ref{enx}).  
Indeed for $a<1$, if we choose to average over the stationary 
distribution, $p_0(X)={1\over {\sqrt {2 \pi D'}}}e^{-X^2/2D'}$, 
then by multiplying both sides of equation (\ref{gnx}) by $p_0(X)$ 
and integrating over $x$ from $-\infty$ to $\infty$, we get after 
straightforward algebra 

\begin{equation}
g_{n+1}=g_n + \sqrt{ {2(1-a^2)}\over {\pi}}\int_0^{\infty}
E_n(y)\, e^{-D'(1-a^2)y^2/2} {\rm erfc}\left( {{ay}\over
{\sqrt 2}}\right)dy .
\label{gn}
\end{equation}
Hence the variance of the number of crossings,
$\sigma_n^2={\langle m^2\rangle}_n -{\langle m\rangle}_n^2$
is given by,
\begin{equation}
\sigma_n^2= g_n + n\lambda -n^2\lambda^2,
\label{var1}
\end{equation}
where $g_n$ is given by the solution of the recursion
equation (\ref{gn}) and $\lambda={1\over {2}}-{1\over
{\pi}}{\sin}^{-1}(a)$. To determine $g_n$ from equation
(\ref{gn}), we need to know the full function $E_n(x)$.

In the next section, we show that there is an alternative
way to derive an expression for $E_n(x)$ without solving the
integral equation (\ref{enx}).

\section{Alternative Derivation of $E_{\mbox{n}}(x)$} 
\label{Alternative Derivation of $E_n(x)$}

An alternative derivation of $E_n(X)$ can be obtained by
noting the evident relation,

\begin{eqnarray}
E_{n+1}(X)-E_n(X)&=& {1\over {2}}\left[ 1-\langle {\rm
sign}(X_n){\rm sign}(X_{n+1})\rangle \right]\nonumber \\ &=&
\langle \theta(X_n)\rangle +\langle \theta(X_{n+1})\rangle -
2 \langle \theta(X_n)\theta(X_{n+1})\rangle,
\label{enxa1}
\end{eqnarray}

where $\theta(x)$ is the standard theta function. Noting
that $X_n$ and $X_{n+1}$ are Gaussian variables, we can
compute the right hand side of equation (\ref{enxa1})
exactly. For this we first need the joint distribution
$P[X_n,X_{n+1}]$ of $X_n$ and $X_{n+1}$ which involves the
correlation matrix $C_{n,n+1}$ given by
\begin{displaymath}
C_{n,n+1}=D'
\left[ \begin{array}{cc}
1-a^{2n} & a(1-a^{2n}) \\
a(1-a^{2n}) & 1-a^{2n+2}
\end{array}\right]
\end{displaymath}
where we have used equation (\ref{nnc}) for the matrix elements.
Using this joint distribution and carrying out the Gaussian
integrations, we get after lengthy but straightforward
algebra the final expression of the right hand side of
equation (\ref{enxa1}) as,

\begin{eqnarray}
E_{n+1}(X)-E_n(X)&=& {1\over {2}}{\rm erfc}\left (-A_n
X\right ) +{1\over {2}}{\rm erfc}\left (-A_{n+1} X\right
)\nonumber \\ &-& {1\over {\sqrt
\pi}}\int_{-A_{n+1}X}^{\infty} dy e^{-y^2} {\rm erfc}\left[
-\sqrt{ {1-a^{2n+2}}\over {1-a^2}} A_n X -a \sqrt{
{1-a^{2n}}\over {1-a^2}}y\right],
\label{enxa2}
\end{eqnarray}
where $A_n = a^n/{\sqrt{2D' (1-a^{2 n})}}$. Changing to
rescaled variable, $x=X/{\sqrt{D'(1-a^2)}}$ and using the
initial condition $E_0(x)=0$, we get from equation
(\ref{enxa2}),

\begin{eqnarray}
E_n(x) &=& {1\over 2}\sum_{m=0}^{n-1} \left[ {\rm erfc}
\left ( -{ a^m x \over \sqrt 2} \sqrt{ 1-a^2 \over 1-a^{2 m}
} \right) + {\rm erfc} \left ( -{ a^{m+1} x \over \sqrt 2}
\sqrt{ 1-a^2 \over 1-a^{2 m +2} } \right) \right. \nonumber
\\ &-& \left.  {2\over \sqrt \pi}\int_{-{a^{m+1} x \over
\sqrt 2} \sqrt{1-a^2 \over 1-a^{2m+2}}}^{\infty} dy e^{-y^2}
{\rm erfc} \left( - \sqrt { 1-a^{2m+2}\over 1-a^2} \sqrt{
1-a^2 \over 1-a^{2 m}} { a^m\over \sqrt 2}x -a \sqrt {
1-a^{2m}\over 1-a^2} y \right) \right] .
\label{enxa3}
\end{eqnarray}

\begin{figure}
\narrowtext \centerline{\epsfxsize\columnwidth
\epsfbox{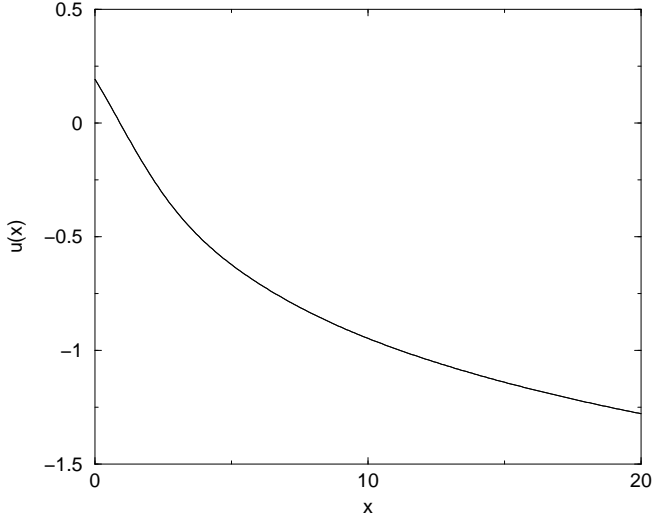}} 
\caption{ Plot of
$u(x)=E_n(x)-n \lambda$ for $a=1/2$ and $n=1000$, calculated
using equation (\ref{enxa3}).  $u(x)$ is symmetric about
$x=0$ .  Also shown is $u(x)$ for $a=1/2$ derived using the
matrix method (equation(\ref{uofxeqn})), the two curves are
indistinguishable.  }
\label{fig100}
\end{figure}

This can be solved numerically, figure (\ref{fig100}) shows $E_n(x)-n \lambda$ for $a=1/2$ and $n=1000$.  
  \widetext

In the next section we discuss the exact asymptotic
properties of the mean and the variance of the number of
sign changes for the two cases: (A) Stable potential,
$\mu>0$, i.e $0<a<1$ and (B) unstable potential, $\mu<0$,
i.e. $a>1$.

\section{Asymptotics of the Mean and Variance of the Number of 
Sign Changes} 
\label{The Exact Asymptotic Properties of the Mean and the Variance of 
Sign Changes}

\subsection{Stable potential: $\mu >0$}

In this case $a=e^{-\mu \Delta\!T}<1$ and also $D'>0$. Thus
as $n\to \infty$, $A_n = a^n/{\sqrt{2D' (1-a^2)}}\to
0$. Then from equation (\ref{enxa2}), we find
$E_{n+1}(X)-E_n(X)\to \lambda$ with $\lambda={1\over
{2}}-{1\over {\pi}}{\sin}^{-1}(a)$ indicating that in the
$n\to \infty$ limit, $E_n(X)$ becomes independent of $X$ (as
long as as $X<a^n$) and to leading order for large $n$,

\begin{equation}
E_n(x) = n\lambda,
\label{en}
\end{equation}
in agreement with equation (\ref{avgm1}). Indeed the
expression for $E_n(x)$ in equation (\ref{enxa3}) is a
solution of the integral equation (\ref{enx}). For finite
$n$, one can write $E_n(x)=n\lambda + u_n(x)$.  An explicit
expression for $u_n(x)$ can be obtained from that of
$E_n(x)$ in equation (\ref{enxa3}). Using the explicit value
of $\lambda$ and after a few steps of algebra we get,

\begin{eqnarray}
u_n(x) &=& {1\over {2}}\sum_{m=0}^{n-1} s_m(x) 
\nonumber \\ s_m(x) &=& {\rm erfc}\left ( -{ a^m x \over
\sqrt 2} \sqrt{1-a^2 \over 1-a^{2m} } \right) +{\rm erfc}
\left ( -{ a^{m+1} x \over \sqrt 2} \sqrt{1-a^2 \over
1-a^{2m+2}} \right)-2 \nonumber +{2 \over \sqrt
\pi}\int_0^{\infty}dy e^{-y^2}{\rm erfc}\left({-{ay}\over
{\sqrt {1-a^2}}}\right) \\ &-&{2 \over \sqrt \pi}\int_{-{x
a^{m+1} \over \sqrt 2} \sqrt{{1-a^{2m} \over
1-a^{2m+2}}}}^{\infty} dy e^{-y^2} {\rm erfc} \left(- \sqrt
{ 1-a^{2m+2} \over 1-a^2} \sqrt{1-a^2 \over 1-a^{2m}} { a^m
\over \sqrt 2}x -a \sqrt { {1-a^{2m}}\over {1-a^2}}y\right).
\label{unxa}
\end{eqnarray}
Besides, using $E_n(x)=n\lambda + u_n(x)$ in
the integral equation (\ref{enx}), we find that $u_n(x)$ also 
satisfies the following integral equation,

\begin{equation}
u_{n+1}(x)={1\over {\sqrt {2\pi}}}\int_0^{\infty}u_n(y)
\left[ e^{-(y-ax)^2/2} + e^{-(y+ax)^2/2}\right]dy + {1\over
{2}} {\rm erfc}\left ( { {ax}\over {\sqrt
2}}\right)-\lambda,
\label{unx}
\end{equation}                                                                     
with $\lambda={1\over {2}}-{1\over {\pi}}{\sin}^{-1}(a)$. 

Now the leading term in $E_n(x)$ is $n\lambda$ and is
independent of $x$ as long as $x<a^{-n}$. This upper cut-off
tends to $\infty$ as $n\to \infty$ since $a<1$.  The $x$
dependence of $E_n(x)$ appears only in the subleading term
$u_n(x)$.  Now, as $n\to \infty$, $u_n(x)$ tends to a
stationary solution independent of $n$ (as long as 
$x \ll a^{-n}$) and is given by the fixed point solution 
$u(x)$ of the integral equation (\ref{unx}),

\begin{equation}
u(x)={1\over {\sqrt {2\pi}}}\int_0^{\infty}u(y) \left[
e^{-(y-ax)^2/2} + e^{-(y+ax)^2/2}\right]dy + {1\over {2}}
{\rm erfc}\left ( { {ax}\over {\sqrt 2}}\right)-\lambda .
\label{ux}
\end{equation}  

We can find $u_n(x)$ perturbatively by expanding the $axy$
term in the exponentials in equation (\ref{ux}) to get,

\begin{equation}
u(x)={1\over {\sqrt {2\pi}}} \sum_{m=0\ ({\rm m
 even})}^{\infty} {(a x)^{m} e^{-(ax)^2/2} \over m!}
 \int_0^{\infty} u(y) e^{-y^2/2} 2 y^m dy + {1\over {2}}
 {\rm erfc}\left( { {ax}\over {\sqrt 2}}\right)-\lambda.
\label{dummy7}
\end{equation}
Defining

\begin{equation}
I_l=\int_0^{\infty} u(x) { a^{l/2} x^l \over \sqrt{l!} }
e^{-x^2/2} dx
\label{dummy8}
\end{equation}
gives

\begin{equation}
u(x)=\sqrt{2 \over \pi} \sum_{m=0\ ({\rm m even})}^{\infty}
{a^{m/2} x^m e^{-(ax)^2/2}\over \sqrt{ m! }} I_m + {1\over
{2}} {\rm erfc}\left ( { {ax}\over {\sqrt 2}}\right)
-\lambda .
\label{uofxeqn}
\end{equation}
Multiplying both sides by $a^{l/2} x^l e^{-x^2/2} /
\sqrt{l!}$ and integrating over positive x gives,

\begin{equation}
I_l= \sum_{m=0\ ({\rm m even})}^{\infty} M_{lm} I_{m} +J_l
\label{miter}
\end{equation}
where

\begin{eqnarray}
M_{lm} ={1 \over \sqrt{2 \pi a}} \left({2 a \over
1+a^2}\right)^{(l+m+1)/2} {\Gamma[(l+m+1)/2] \over \sqrt{l!
m!}}  {\rm \,\,\,\,\, m \,\, even, \,\, 0 \,\, otherwise} \\
J_l={a^{l/2} \over \sqrt{l!}} \int_0^{\infty} x^l e^{-x^2/2}
\left[{{1\over2} \rm erfc}\left ( { {ax}\over {\sqrt
2}}\right) -\lambda \right].
\end{eqnarray}
Thus

\begin{equation}
{\bf I} = ({\bf 1}-{\bf M})^{-1} {\bf J}
\label{dummy9}
\end{equation}
This perturbative expansion in $a$ is the matrix method.
Although $M_{lm}$ is an infinite array, the elements
decrease rapidly with increasing $l$ and $m$ since each 
increment of $l$ and $m$
gives a higher power of $a$ ($a<1$).  We can solve this
numerically. Figure \ref{fig100} shows $u(x)$ for $a=1/2$.
Alternatively, using {\sc mathematica}, we can solve
equation (\ref{miter}) iteratively.  At each iteration we
obtain a new ${\bf I}$ whose elements are a series in $a$ up
to our required order.  Convergence is rapid.  This can only
be done with a smaller matrix than the numerical method, but
gives a result for general $a$.  Note that we could
equivalently have done this perturbative expansion on
equation (\ref{enx}) to get $E(x)$ and subtracted the $n
\lambda$.

Alternatively, we can directly take the $n \to \infty$ limit
of the expression of $u_n(x)$ in equation (\ref{unxa}) to
get,

\begin{equation}
u(x)={1 \over 2} \sum_{m=0}^{\infty} s_m(x),
\label{ux1} 
\end{equation}
where $s_m(x)$ is given by equation (\ref{unxa}).  This has
been done numerically (for $n$ large but finite) and found
to agree with the matrix method.

Let us first compute the asymptotic properties of the fixed
point solution $u(x)$.  Consider first the limit $x\to 0$.
Putting $x=0$ in equation (\ref{unxa}) and carrying out the
integrations we get, after some algebra,

\begin{equation}
u(0)={1\over {\pi}}\sum_{m=0}^{\infty} {\sin}^{-1}\left[ {
{a\sqrt {1-a^2}}\over {\sqrt {1-a^{2m+2}}} } \left(
1-\sqrt{1-a^{2m}}\right) \right].
\label{u01}
\end{equation}
For example, for $a=1/2$, we get $u(0)=0.19160374\ldots$
which agrees very well with the result obtained from the
direct numerical integration of equation (\ref{unx}) in the
large $n$ limit. Consider now the other limit $x\to
\infty$. By making the change of variable, $y-ax ={\sqrt
2}z$ in the integral equation (\ref{ux}), we get to leading
order for large $x$ (where the lower limit of the first
integration $\to -\infty$),

\begin{equation}
u(x)\approx u(ax) -\lambda .
\label{uxas}
\end{equation}
The solution of this equation is given by,

\begin{equation}
u(x)\approx {\lambda \over {\ln a}}{\ln x}.
\label{uxas1}
\end{equation}
Note that $\lambda/{\ln a}<0$ for $a<1$ and hence $u(x)$
goes to $-\infty$ logarithmically as $x\to \infty$.  This is
consistent with the fact that $E_n(x)\approx n\lambda +
\lambda{ {\ln x}\over {\ln a}} \sim { {\lambda}\over {\ln
a}}{\ln (xa^n)}\to 0$ as $x\to a^{-n}$ as it should
evidently from the direct expression of $E_n(x)$ in equation
(\ref{enxa3}).

In figure \ref{fig100}, we plot $u(x)$ obtained from
numerically evaluating the sum in equation (\ref{unxa}) and
also the result obtained using the matrix method.  The
results agree to within numerical precision.  Also, they
agree with the asymptotic results in the large and small $x$
limits.  Note that the $x \to \infty$ limit was not
attainable by the matrix or summation methods because in
both cases the evaluation is done to a finite order or
finite $n$.

\begin{figure}
\narrowtext \centerline{\epsfxsize\columnwidth
\epsfbox{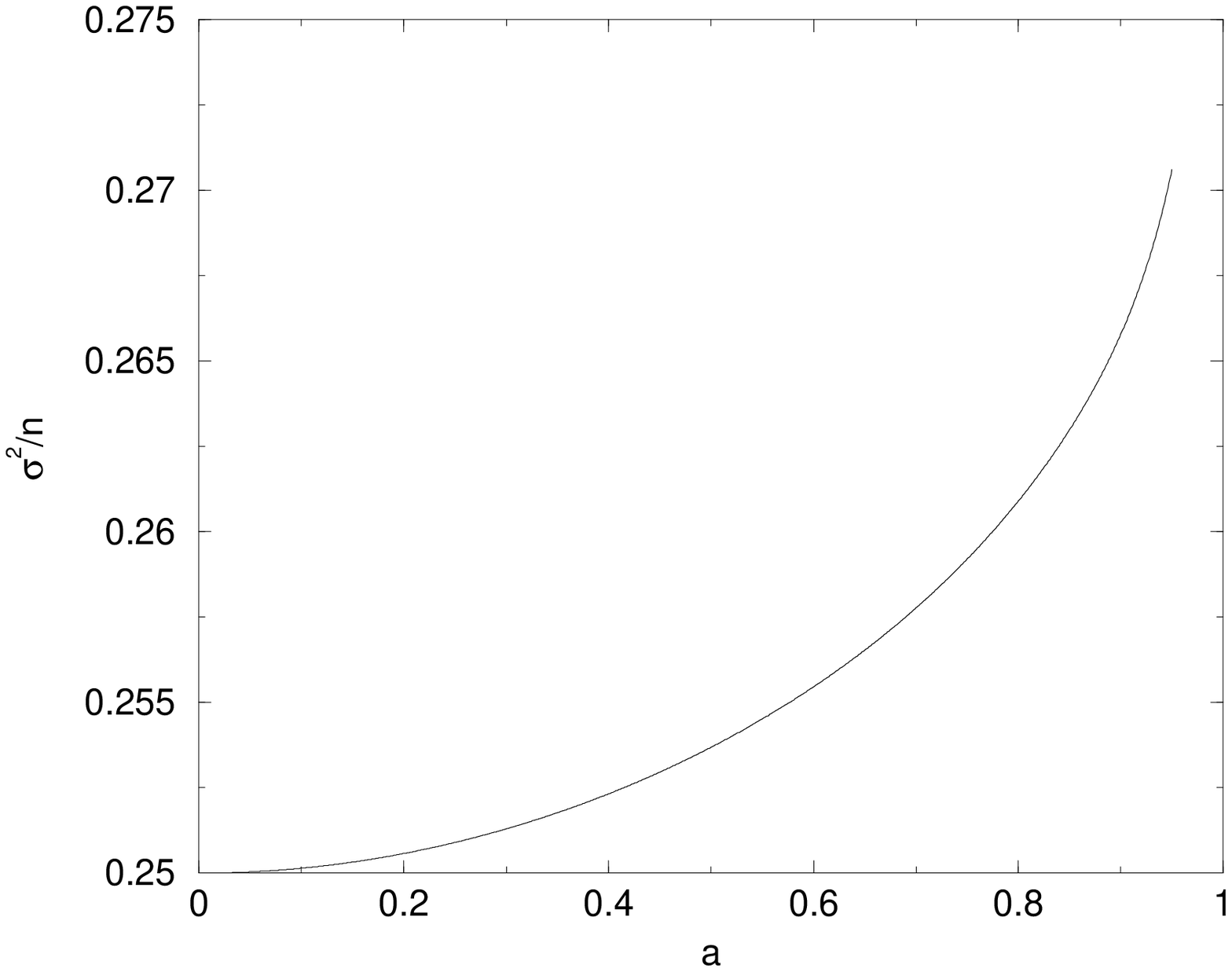}} \caption{Plot of $\sigma_n^2
/n$ against $a=e^{-\mu \Delta\!T}$ with $D'=1$. Note that 
$\sigma_n^2/n \to \infty$ for $a \to 1$, and the series has not
converged for large $a$.}
\label{fig600}

\end{figure}

Once we know $u_n(x)$, then we can determine the variance
$\sigma_n^2$ from equation (\ref{var1}). Substituting
$E_n(x)=n\lambda + u_n(x)$ in equation (\ref{gn}) and using
the fact that $u_n(x)$ tends to the fixed point solution
$u_n(x)\to u(x)$ for large $n$, we get

\begin{equation}
g_{n+1} = g_n + 2n \lambda^2 +\beta,
\label{gn1}
\end{equation}
where

\begin{equation}
\beta=\sqrt{ {2(1-a^2)}\over {\pi}}\int_0^{\infty} u(y)\,
e^{-D'(1-a^2)y^2/2} {\rm erfc}\left( {{ay}\over {\sqrt
2}}\right)dy.
\label{beta}
\end{equation}
One can easily solve the recursion equation (\ref{gn1})
exactly using $g_0=0$ and we get,

\begin{equation}
g_n= \lambda^2n^2 + (\beta-\lambda^2)n,
\label{gn2}
\end{equation}
where $\beta$ is given by equation
(\ref{beta}). Substituting this expression for $g_n$ in
equation (\ref{var1}), we finally get the required exact
expression of the variance for large $n$,

\begin{equation}
\sigma_n^2= \left( \lambda-\lambda^2 +\beta\right) n .
\label{var2}
\end{equation}
Thus the variance can be exactly determined once we know the
function $u(x)$ and thereby $\beta$ from equation
(\ref{beta}). Using the exact expression of $u(x)$ from
equation (\ref{ux1}), we have, in principle, an exact result
for $\beta$ and hence for $\sigma_n^2$.  Substituting the
$u(x)$ derived from the matrix method into equation
(\ref{beta}) gives $\beta$ and hence $\sigma_n^2$.
$\sigma_n^2 /n$ is plotted as a function of $a$ for $D'=1$
in figure (\ref{fig600}).

  \widetext

\subsection{ Unstable potential: $\mu<0$}

\begin{figure}
\narrowtext \centerline{\epsfxsize\columnwidth
\epsfbox{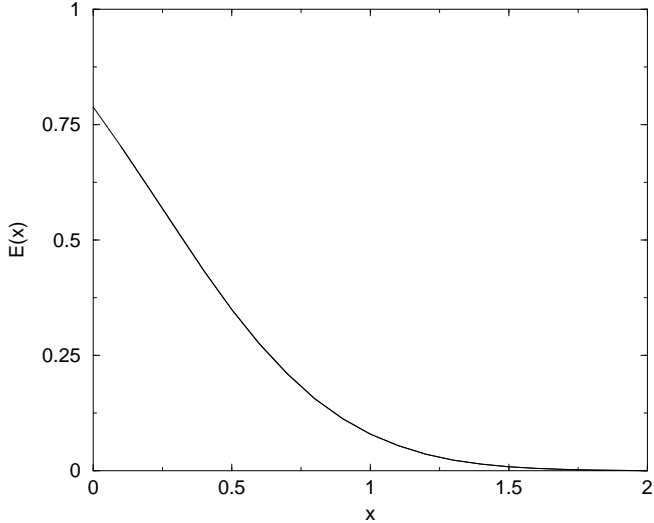}} \caption{The
expected number of detected crossings $E(x)$ for the case
$a=2$ (unstable potential).  Shown are the results of the
matrix method and also numerical evaluation of equation
(\ref{exa3}) which are indistinguishable. }
\label{fig500}

\end{figure}

In the previous subsection, we have seen that for $a<1$, the
function $E_n(x)$ behaves asymptotically for large $n$ as
$E_n(x)=n\lambda+u(x)$ where $u(x)$ is given either by the
exact expression in equation (\ref{ux1}) or equivalently by
the solution of the integral equation (\ref{ux}). For the
unstable potential $\mu<0$, i.e. $a>1$, the number of
crossings will be finite and so $E_n(x)$ approaches a steady
state as $n\to \infty$.  This is most easily seen from
equation (\ref{enxa2}). For $a>1$, $A_n=a^n/{\sqrt{2D'
(1-a^2)}}\to \infty$ as $n\to \infty$ (note that $D'=D/\mu
<0$). Taking this limit in equation (\ref{enxa2}), we see
that $E_{n+1}(x)-E_n(x)\to 0$ for all $x$ as $n\to \infty$,
indicating $E_n(x)\to E(x)$ as $n\to \infty$. This steady
state $E(x)$ is given by the fixed point solution of the
integral equation (\ref{enx}) with $a>1$,

\begin{equation}
E(x)={1\over     {\sqrt    {2\pi}}}\int_0^{\infty}E(y)
\left[ e^{-(y-ax)^2/2} +  e^{-(y+ax)^2/2}\right]dy +
{1\over {2}} {\rm erfc}\left ( { {ax}\over {\sqrt 2}}\right).
\label{enx1}
\end{equation}                                            
$E(x)$ can be found using the matrix method in the same way
as before but with $J_l$ replaced by $J_l'$ where

\begin{eqnarray}
J_l'={a^{l/2} \over \sqrt{l!}} \int_0^{\infty} x^l
e^{-x^2/2} {{1\over2} \rm erfc}\left ( { {ax}\over {\sqrt
2}}\right).
\end{eqnarray}
$E(x)$ is shown in figure (\ref{fig500}) for the case $a=2$.

  \widetext

Alternatively, $E(x)$ is also given by taking the $n\to
\infty$ limit of the exact expression in equation
(\ref{enxa3}) (with $a>1$),

\begin{eqnarray}
E(x) &=& {1\over {2}}\sum_{m=0}^{\infty}\left[ {\rm erfc}
 \left ( -{ {a^m x}\over {\sqrt 2}} \sqrt{1-a^2 \over
 1-a^{2m} } \right) + {\rm erfc} \left ( -{ {a^{m+1} x}
 \over {\sqrt 2}} \sqrt{1-a^2 \over 1-a^{2m+2} } \right)
 \right. \nonumber \\ &-& \left.  {2\over {\sqrt
 \pi}}\int_{-{a^{m+1} x \over \sqrt 2} \sqrt{1-a^{2m} \over
 1-a^{2m+2}}}^{\infty} dy e^{-y^2} {\rm erfc} \left(- \sqrt
 { 1-a^{2m+2} \over 1-a^2 } \sqrt { 1-a^2 \over 1-a^{2m} } {
 a^m \over \sqrt 2}x -a \sqrt { {1-a^{2m}}\over
 {1-a^2}}y\right)\right].
\label{exa3}
\end{eqnarray}
This is shown in figure (\ref{fig500}). 

We have calculated the variance of the number of detected
crossings and also the expected number of detected crossings
starting at position $x$.  These calculations have been carried 
out by two independent methods and the results agree with other.

\section{ Partial Survival} \label{Partial Survival}

\begin{figure}
\narrowtext \centerline{\epsfxsize\columnwidth
\epsfbox{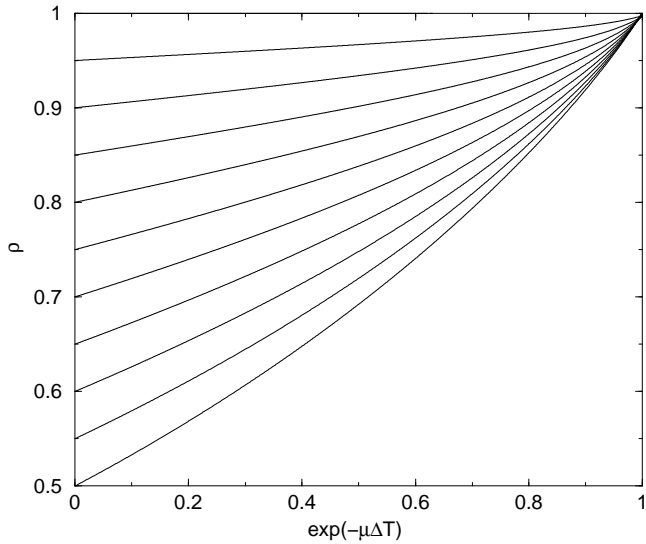}} \caption{
Plot of the random walk discrete persistence eigenvalue
$\rho_p(a)$ for partial survival against $a=e^{-\mu
\Delta\!T}$ for values of the survival probability $p$ from
0 (normal discrete persistence, lowest curve) to 1
(guaranteed to survive so $\rho_1(a)=1$, top curve) in steps
of 0.1.  The curves are the raw series in $a$ to order
$a^{50}$.  Note that for all the curves, $\rho_p \to 1$ for
$a \to 1$ since a walker will always survive for a time
$\Delta\!T$ when $\Delta\!T \to 0$. Since the series in $a$
are finite, they do not quite converge to 1 in this limit}
\label{fig2001dx}

\end{figure}

\begin{figure}
\narrowtext \centerline{\epsfxsize\columnwidth
\epsfbox{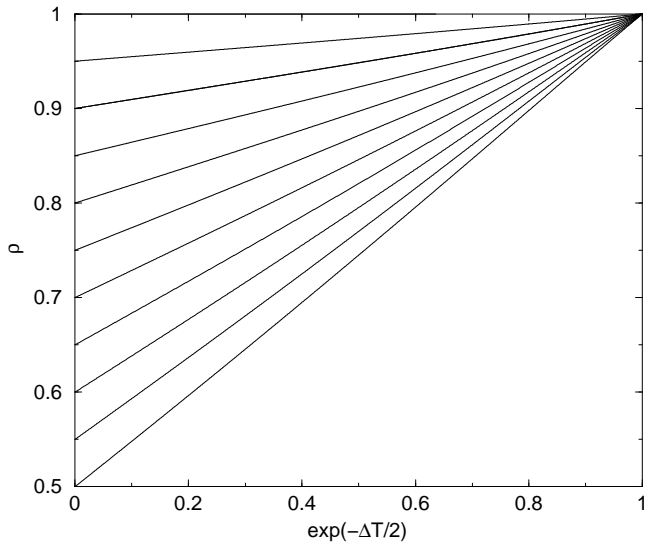}}
\caption{ Plot of the random acceleration discrete
persistence eigenvalue $\rho_p(a)$ for partial survival
against $a=e^{-\Delta\!T/2}$ for values of $p$ from 0
(normal discrete persistence, lowest curve) to 1 (guaranteed
to survive so $\rho_1(a)=1$, top curve) in steps of 0.1.
The curves are Pad\'es of the raw series in $a$ to order
$a^{19}$.  }
\label{fig2001ddx}

\end{figure}

\begin{figure}
\narrowtext \centerline{\epsfxsize\columnwidth
\epsfbox{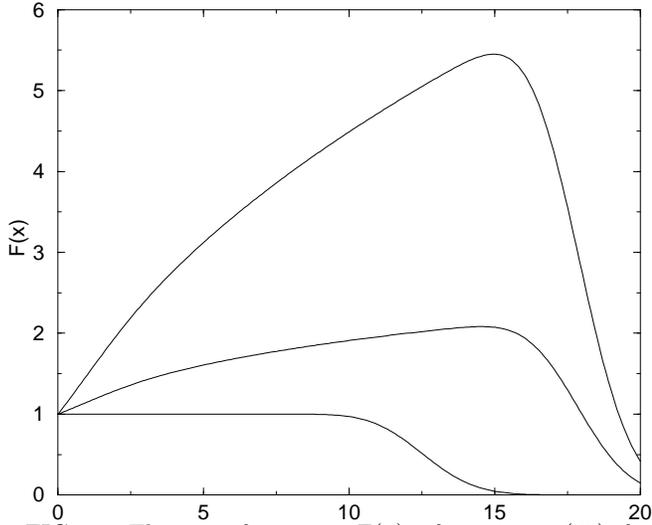}}
\caption{The eigenfunctions $F(x)$ of equation (\ref{ps1})
for $a=0.5$ with $p=1$ (lowest curve), $p=0.5$ (middle
curve) and $p=0$ top curve.  The eigenfunctions are defined
only up to an arbitrary prefactor which has been chosen here
so that $F(0)=1$.  Since the eigenfunctions are series in
$x$, for large $x$ they do not converge to the correct
solution.  This can be clearly seen for the $p=1$ case
where, since the walker is guaranteed to survive, $F(x)$ is
a constant everywhere whereas the plot is not constant for
$x \ge 10$.  For the $p=0$ case, $F(x) \sim x^\nu$ with $\nu
=\ln \rho / \ln a$. }
\label{fig2011}

\end{figure}

The partial-survival probability, $F_n(p,x)$, is the
probability of surviving beyond the $n$th sampling having
started at $x$ if each detected crossing of the origin is
survived with probability $p$.  Thus,

\begin{equation}
F_n(p,x)=\sum_{m=0}^{\infty}Q_n(m,x)p^m
\label{dummy10}  
\end{equation}
and as stated before $F_n(p,x)$ is the generating function
for $Q_n(m,x)$.  $F_n(p,x)$ satisfies the integral equation
(\ref{rec2}).  We expect that for large $n$,
$F_n(p,x)=[\rho_p(a)]^n F(x)$ where $\rho_p(a)=e^{-\theta(p)
\Delta\!T}$ and $\theta(p)$ is the discrete persistence
exponent.  Substituting this into equation (\ref{rec2}), we
get an eigenvalue equation for $F(x)$,

\begin{equation}
\rho_p(a) F(x)= {1\over {\sqrt {2\pi}}}\int_0^{\infty} F(y)
\left[ e^{-(y-ax)^2/2} + p e^{-(y+ax)^2/2}\right] dy.
\label{ps1}
\end{equation}
The largest eigenvalue $\rho_p(a)$ and the corresponding
eigenfunction can then be determined either by the matrix
method or by the variational method as in our previous paper
\cite{mbe}.  Using the matrix method, we get,

\begin{eqnarray}
\rho_p(a) F(x) &=& { e^{-a^2 x^2 /2} \over \sqrt{2 \pi}}
\sum_{m=0}^{\infty} {a^m \over m!} x^m (1+(-1)^m p) I_m
\label{eqnno39} \\ \rho_p(a) I_l &=& \sum_{m=0}^{\infty}
\left( G_{lm}+p G_{lm}' \right) I_m
\label{ps2}
\end{eqnarray}
where 

\begin{eqnarray}
I_m &=& {a^m \over m!} \int_0^{\infty} dy y^m e^{-y^2/2}
F(y) \\
\label{dummy11} G_{lm} &=& { 1\over \sqrt{8 \pi} } a^m 
\left({ 2 \over 1+a^2 }\right)^{(l+m+1)/2}
{\Gamma[(l+m+1)/2] \over {l! m!} }
\label{dummy12}
\end{eqnarray}
and $G_{lm}'(a)=G_{lm}(-a)$.  In fact, $G_{lm}$ is the
matrix used for calculating the discrete persistence
exponent \cite{mbe}, whilst $G_{lm}'$ gives alternating
persistence.  This is to be expected, since for $p=0$ we
just have ordinary persistence and for $p \gg 1$ we would
expect the paths which cross between every sampling
(alternating persistence) to dominate.  In the same way, one
may calculate $\rho_p(a)$ for the discretely sampled random
acceleration problem studied in \cite{ebm}, whose stationary
process is given by

\begin{equation}
\ddot X+(\alpha+\beta) \dot X +\alpha \beta X = \eta(T)
\label{dummy13}
\end{equation}
where $\eta(T)$ is Gaussian white noise with mean zero and
correlator $<\eta(T)\eta(T')>=2 \alpha \beta (alpha +\beta) \delta(T-T')$, and
$\alpha=1/2$, $\beta=3/2$ for the random acceleration
problem, although other values of $\alpha$, $\beta$ can be
considered.  We get

\begin{equation}
\rho_p(a) I_{ij} = \sum_{k,l=0}^{\infty} 
\left( H_{ijkl}+p H_{ijkl}' \right) I_{kl}
\label{dummy14}
\end{equation}
where $H_{ijkl}$ and $I_{ij}$ are given in \cite{ebm}.
Again, $H_{ijkl}$ is the matrix used to find the discrete
persistence exponent and $H_{ijkl}'(a)=H_{ijkl}(-a)$ gives
the alternating persistence exponent.  We find $\rho_p(a)$
numerically and also as a power series in $a$ for the two
processes given above.  The results are shown in figures 
(\ref{fig2001dx},\ref{fig2001ddx}).  Also the eigenfunction $F(x)$ for the
Ornstein-Uhlenbeck Process may be found by substituting the
eigenvector corresponding to the largest eigenvalue into
equation (\ref{eqnno39}). Results for $a=0.5$ with $p=0$,
$0.5$, $1$ are shown in figure (\ref{fig2011}).

  \widetext

  \widetext

  \widetext

${}$ \newline
  
So  far we  have found  the  mean and  variance and  also the  
partial-survival  probability  for the  Ornstein-Uhlenbeck  Process, a  
simple Gaussian  Markov problem. We did this by using the  propagator,
$P(Y,\Delta\!T|X,0)$ (equation(\ref{propagator})).  We also showed how
the partial-survival probability of other GSPs of known propagator can
be found by  using the perturbative matrix method,  and we illustrated
this for  the random acceleration problem.  However,  the methods used
become progressively harder as the  number of variables in the problem
increases.  The Ornstein-Uhlenbeck Process  had only the position $X$,
the  random  acceleration problem  had  $X$  and  $V$, while  the 
persistence problem for the  diffusion equation cannot be expressed in
terms  of a  propagator with  a finite  number of  variables.   In the
remainder of this paper we will  use a different approach based on the
correlator  of  the process  $C(T)$.   This  removes the  difficulties
mentioned above and  can, furthermore, be applied to  any GSP of known
correlator.  The  results for low-dimensional  problems obtained above
are slightly more  accurate than those given by  the correlator method
because they  can be calculated to  higher order. They can  be used as
powerful checks of the accuracy of the correlator method.

\newpage

\part{The Correlator Expansion} \label{The Correlator Expansion}

The correlator expansion was  initially used to calculate the discrete
persistence exponent of an arbitrary Gaussian Stationary Process (GSP)
and  also,  through  extrapolation  to the  continuum,  the  continuum
persistence exponent  \cite{ctl}.  Here we  will extend the  method to
calculate  the occupation-time distribution  and the  distribution of
crossings.  As  these calculations  will require some  explanation, we
take this opportunity to describe the correlator expansion in full.

\section{An introduction to the correlator expansion} 
\label{An introduction to the correlator expansion}

The expansion starts from the following identity for $P_n$,
the probability of no detected crossings in $n$ samplings:

\begin{equation} 
P_n = \left< \prod_{i=1}^n \Theta [X(i \Delta \! T)] \right>
\label{ii10} 
\end{equation}
where $\Theta(X)$ is the Heaviside step function and the
expectation value is taken in the stationary state.  One may
write $\Theta[X(i \Delta\!T)] =(1+\sigma_i)/2$, where
$\sigma_i \equiv {\rm sign}[X(i\Delta\!T)]$, and expand the
product to give,

\begin{equation} 
P_n ={1\over 2^n} \left(\!1 +\!  \sum_{1=i<j}^n
\left<\sigma_i \sigma_j \right>+\sum_{1=i<j<k<l}^n
\left<\sigma_i \sigma_j \sigma_k \sigma_l \right> +\ldots
\right)
\label{ii20} 
\end{equation}
where the terms with odd numbers of $\sigma$s vanish since the 
process is  symmetric under  $X \to  -X$ (and therefore under 
$\sigma \to -\sigma$. To evaluate the terms we use the representation

\begin{equation} 
\sigma_l= {1\over i \pi} \lim_{\epsilon \to 0}
\int_{-\infty}^\infty {dz_l\,z_l\, e^{i z_l X_l} \over
(z_l-i \epsilon) (z_l+i \epsilon)}
\label{ii30}
\end{equation} 
Carrying out the required averages of the  Gaussian process gives 
the correlation functions appearing in (\ref{ii20}):

\begin{equation}
\left <\sigma_{l_1} \dots
\sigma_{l_m}\right>=\int\prod_{j=1}^m\left({dz_j \over i \pi
z_j}\right)\exp\left(-{1\over
2}z_{\alpha}\,C_{\alpha\beta}\, z_{\beta}\right),
\label{ii40}
\end{equation}
where $C_{\alpha\beta}= \langle
X[\alpha\Delta\!T]\,X[\beta\Delta\!T] \rangle =
C(|\alpha-\beta|\Delta T)$, and there is an implied
summation over $\alpha$ and $\beta$ from $1$ to $m$. Notice
that we have already taken the limit $\epsilon \to 0$ in
(\ref{ii40}), with the understanding that all integrals are
now principal part integrals.

For the $m=2$ case this integral can be done exactly by
differentiating with respect to $C_{12}$ and doing the two
simple Gaussian integrals before integrating again with
respect to $C_{12}$ and imposing the boundary condition
that $\left<\sigma_{l_1} \sigma_{l_2}\right> =0$ for $C_{12}=0$.  
This gives the well-known result $\left<\sigma_{l_1} \sigma_{l_2}\right> 
= (2/\pi) \sin^{-1}C_{12}$.  For $m \ge 4$ this method becomes 
non-trivial.  Instead, we choose to expand the exponential in
equation (\ref{ii40}) in powers of $C_{\alpha\beta}$
($\alpha \ne \beta$) leaving the terms with $\alpha=\beta$
unexpanded (noting that $C_{\alpha\alpha}=1$).  This allows
us to evaluate each correlation function of the $\sigma s$ up 
to a given order in the correlators $C_{\alpha\beta}$.  
By symmetry, only terms which generate
odd powers of every $z_{\alpha}$ in the expansion of the
exponential (to give even powers overall in the integrand,
through the factors $1/z_i$) give a non-zero integral.  This
suggests a simple diagrammatic representation for the terms
in (\ref{ii20}), as given by (\ref{ii40}).  On a
one-dimensional lattice containing $n$ sites, with lattice
spacing $\Delta\!T$, draw $m$ vertices at the locations
$l_1,l_2, \dots,l_m$.  Connect the vertices by lines in all
possible ways (summing over these different possibilities)
subject to the constraint that each vertex is connected to an odd 
number of lines. Associate a factor $\sqrt{2 \pi} (p-2)!!$ (coming 
from evaluating the Gaussian integrals) with each vertex of order
$p$, a factor $(-C_{l_il_j})^r/r!$ with the $r$ lines
connecting site $l_i$ to site $l_j$, and an overall factor
$(\pi i)^{-m}$ with the diagram.  This suffices to evaluate
the integrals in (\ref{ii40}).  Evaluating the summations in
(\ref{ii20}) involves enumerating all configurations of the
vertices on the lattice for a given ordering of the points,
and noting that the factor $C_{l_il_j}$ associated with a
given line is equal to $C(q\Delta\!T)$, where $q=|l_i-l_j|$
is the length of the line in units of $\Delta\!T$.

We choose to define $C(\Delta\!T)$ as $1^{st}$ order small
and $C(q\Delta\!T)$ as $q^{th}$ order small.  Although
somewhat arbitrary, this is a large $\Delta\!T$ expansion
and most processes of physical interest have correlators
$C(q\Delta\!T)$ which decrease exponentially for large
argument, so our definition is consistent for large 
$\Delta\!T$. For the random walk, the correlator is 
$C(T)=e^{-\Delta\!T/2}$ and the definition is always valid.
For other processes it is often possible to re-expand the
correlator in terms of an exponential and work to a given 
order (in practice we can go up to 14th order) in this 
exponential.  This can be done for, e.g., the random 
acceleration problem.  The order of a diagram is then
equal to the total length of its lines (measured in units of
the lattice spacing $\Delta\!T$).  Thus to a given order $k$, 
we need only evaluate correlations functions with separations 
up to $2k\,\Delta T$. 

\begin{figure} 
\narrowtext \centerline{\epsfxsize\columnwidth
\epsfbox{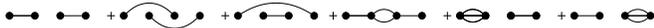}}
\caption{All topologically distinct contributions to
$\langle \sigma_i\sigma_j\sigma_k\sigma_l\rangle$ up to
fourth order.  Note that the 1st, 5th and 6th diagrams are
disconnected, whilst the others (including diagrams 2 and 3)
are connected (due to the constraint that the order of the
points $i,j,k,l$ must be unchanged). }
\label{fii10}
\end{figure}

To illustrate this approach, we show in Figure \ref{fii10}
all the topologically distinct diagrams contributing to
$\langle \sigma_i\sigma_j\sigma_k\sigma_l\rangle$ up to
$4^{th}$ order. The first diagram, when enumerated on the
lattice, will be $2^{nd}$ order or greater, while the
remaining five will be $4^{th}$ order or greater.

  \widetext

\begin{figure} 
\narrowtext \centerline{\epsfxsize\columnwidth
\epsfbox{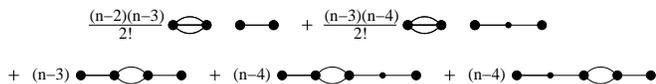}}
\caption{Enumeration of two of the diagrams from figure
\ref{fii10}, with embedding factors, up to fifth order.
Large dots are vertices, small dots intermediate sites.  The
final diagram here gives a contribution $(n-4) (2/ \pi^2)
C(2 \Delta\!T) C(\Delta\!T)^3$ to $P_n$.}
\label{fii20}
\end{figure}

In Figure \ref{fii20} are shown the enumerations of two of
the basic diagrams of figure \ref{fii10} together with their
embedding factors (the number of ways they can be placed on
the lattice), up to 5th order.
  \widetext

Thus the calculation of $P_n$ proceeds in 3 stages:  

1: all the basic diagrams up to the required order are 
found, for example the 4-vertex diagrams shown in figure
\ref{fii10}.

2: the basic diagrams are enumerated on the lattice,
including the `stretched' diagrams, as shown in figure
\ref{fii20}.

3: the appropriate factors are assigned to each diagram.

The total number of enumerated diagrams increases roughly by
a factor of 2 for each extra order. At 14th order there are
12\,434 diagrams.  The process was automated using {\sc
mathematica}.  For calculating $P_n$, finding the basic
diagrams was the most challenging task in terms of computer
time and memory.  To achieve 14th order we used the fact
that all disconnected diagrams can be constructed by
combining two or more connected diagrams.  At 14th order,
diagrams with 12 or more vertices are disconnected, thus
only connected diagrams with up to 10 vertices need be
found.  Furthermore, diagrams containing 2 or more lines
connecting the same points can be constructed from diagrams
with only 0 or 1 lines connecting points by adding pairs of
lines.  The procedure adopted was as follows:

For the 2 to 10-vertex diagrams: all possible diagrams up to
14th order with only 0 or 1 lines connecting points {\it
and} all vertices odd are constructed.  To these diagrams
pairs of lines are added in all possible ways up to 14th
order.  The connected diagrams are then selected and stored.
                                
For 2 to 28-vertex diagrams: all diagrams are constructed by
combining the connected diagrams found above, for example,
the 6-vertex diagrams are formed from 3 2-vertex diagrams,
one 2-vertex and one 4-vertex diagrams, and one 6-vertex
diagram (with the appropriate permutation factors arising from
the various ways of ordering the connected diagrams).  For
large vertex numbers, this produces significant savings over
naively trying all diagrams (since the vast majority of
diagrams do not satisfy the odd-vertex criterion).  In this
way, up to order $k$ we need only find all connected diagrams 
with up to $q$ vertices, where $q=(2k+4)/3$ is even and we
round down, whilst the diagrams go up to $2k$ vertices.

Having found $P_n$, we find $\rho \, (=e^{-\theta
\Delta\!T})$ using the fact that $P_n \sim \rho^n$ for large
$n$. Thus $\rho$ is formally obtained as 
$\rho = \lim_{n \to \infty} P_{n+1}/P_n$. However, since we 
started in the stationary state, the relation $\rho = P_{n+1}/P_n$ 
in fact holds for all $n$ larger than the length of the longest diagram.  
Expanding the expression for $P_{n+1}/P_n$ as a Taylor series up 
to 14th order in $a=e^{-\Delta\!T/2}$ gives us a series expansion 
for $\rho(a)$ about $\Delta\!T=\infty$ ($a=0$).
For the random walk, for example, 
$C(\Delta\!T)=e^{-\Delta\!T/2}=a$ for $\mu=1/2$,
substituting this into our expression for $\rho$ gives us a
series up to 14th order in $a$ whose coefficients agree with
those of the matrix method (to within the numerical error of
the matrix method).

For the usual random acceleration problem, $C(\Delta\!T)=(3
e^{-\Delta\!T/2} -e^{-3 \Delta\!T/2})/2=(3 a-a^3)/2$. For
this case our identification of $C(14 \Delta\!T)$ as being of 
the same order as $C(\Delta\!T)^{14}$ does not hold for all
$\Delta\!T$.  However, if we only keep terms up to $a^{14}$
our expansion be exact up to order 14 in $a$. Whenever possible,  
this is what we will always do.  Note that in this way we are now
working strictly to 14th order, even though $C(j \Delta\!T)
\ne C(\Delta\!T)^j$.

In \cite{mbe} we introduced the concept of alternating
persistence, with $P^{\rm{A}}_n$ being the probability that
$X_i$ is positive for odd $i$ and negative for even $i$ (or
vice versa).  We find $\rho_A$ by noting that, whereas
before we required $X_1,X_2, \dots X_n >0$, we now require
$X_1,-X_2,X_3,-X_4 \dots X_n >0$.  Thus the calculation is
as before except that $C(q \Delta\!T) \to -C(q \Delta\!T)$
for $q$ odd.  Making this minor change to the normal
persistence result gives us the alternating persistence
exponent.  This way of accounting for sign changes between
$X_i$, $X_j$ will be used below to calculate the distribution 
of crossings.

We have applied the correlator expansion to the random walk,
$\dot x = \eta(t)$ and random acceleration, $\ddot x =
\eta(t)$ using the transformation to logarithmic time to generate 
the corresponding stationary processes. Furthermore we have studied 
diffusion from random initial conditions in 1-3 dimensions, 
$\partial \phi/ \partial t = \nabla^2 \phi$,  where 
$\phi({\bf x},t)$ is the diffusion field and the initial condition 
$\phi({\bf x},0)$ is delta-correlated Gaussian noise. We consider 
the persistence of $\phi$ at a single site, for example 
$\phi({\bf 0},t)$.  For this process the correlator is:

\begin{equation} 
C(T)={\rm sech}^{d/2}(T/2)
\label{ii50} 
\end{equation}
where $d$ is the space dimension.  As for the random
acceleration, we define $a=\exp(-\Delta\!T/2)$ for $d=2$ and
$a=\exp(-\Delta\!T/4)$ for $d=1,3$ and then expand the
correlator in powers of $a$.  For $d=1,2$ the lowest power
of $a$ is $a^1$ and so we expand up to $a^{14}$ whilst for
$d=3$ the lowest power is $a^3$ and so we expand up to
$a^{42}$.  We also considered the processes $d^n x / dt^n =
\eta(t)$ for $n>3$.  In logarithmic time the correlators
are \cite{IIA1}:
\begin{equation} 
C_n(T)=(2-1/n) e^{-T/2} {_2F_1} (1,1-n;1+n;e^{-T})
\label{ii60} 
\end{equation}
where ${_2F_1}$ is the standard hypergeometric function.
The $n=1,2$ cases are the random walk and random
acceleration, whilst the limit $n \to \infty$ case 
reproduces the correlator for the $d=2$ diffusion process 
mentioned above \cite{IIA1}.

For all these problems we define a discrete persistence exponent,  
$\theta_D(a) = -\ln\rho(a)/\Delta\!T = \ln\rho(a)/2\ln a$, and 
plot $\theta_D(a)$ against $a$ for $0 \le a \le 1$, i.e.\ 
$\infty \ge \Delta\!T \ge 0$.  Since we are plotting finite series 
in powers $a$, they do not converge for $a \to 1$.  This problem is 
exacerbated by the $1/\ln a$ term in the definition of $\theta_D(a)$, 
which causes $\theta_D(1)$ to blow up unless $\rho(1)=1$.  
To make $\rho$ and hence $\theta_D(a)$ more accurate for $a$ close 
to 1 we extrapolate $\theta_D(a)$ to the continuum.  To do this we
use the technique of Pad\'e approximants borrowed from the
field of series expansions for critical phenomena 
\cite{DombAndGreenVol3}.  The Pad\'e approximant involves
replacing the 14th order series in $a$ with a fraction whose
numerator and denominator are series in $a$.  The sum of the
order of these two series is 14 and the coefficients are
chosen so that when the fraction is expanded as a series in
$a$ it is identical to the raw series. 
This approach markedly improves the results for $\rho$.  
For example, for the random walk, the Pad\'e of the 14th order 
series appears to better the 25th order raw series 
obtained from the matrix method (both raw series agree, 
of course, up to 14th order). However, in order to get accurate
continuum results for $\theta$ we add 1 further term to the
Pad\'e (either numerator or denominator) whose coefficient
is chosen so that the exact constraint $\rho(1)=1$ is 
satisfied. This serves to give reasonably accurate estimates 
of the continuum persistence exponent. For example, for the 
random acceleration problem we find $\theta =0.2506(5)$ from 
the Pad\'e approach, compared to the exact result of $1/4$.

\begin{figure} 
\narrowtext \centerline{\epsfxsize\columnwidth
\epsfbox{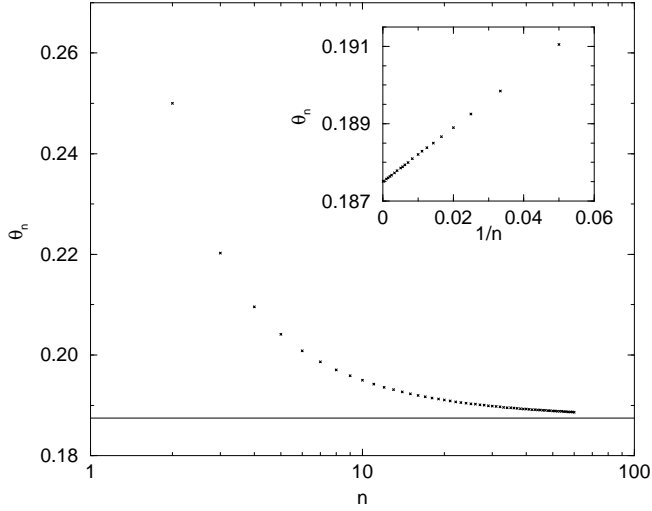}}
\caption{Plot of $\theta_n$ against $n$ for small $n$.  The
$n=1,2$ results are omitted as they are known analytically
to be $1/2$ and $1/4$ respectively and only one constraint
may be imposed on the Pad\'e in these cases.  Note that
$\theta_n$ goes to the continuum result of $0.1875(1)$ (solid
line) rather slowly, in fact as $1/n$, see inset.  The
results were obtained using Pad\'es with 2 constraints, an
average being taken of suitable Pad\'es of order 14 to 10.
Inset: Plot of $\theta_n$ against $1/n$ showing that
$\theta_n -\theta_{\infty} \propto 1/n$ for $n > 20$.}
\label{fdnxplot3to10ANDANDfdnxplotvs1overi}
\end{figure}

For certain sufficiently smooth processes the derivative of
$\theta_D(\Delta\!T)$ at $\Delta\!T=0$ is zero \cite{ebm}.  
We can thus add a further term to the Pad\'e to impose this 
constraint, and markedly improve the accuracy of our estimate 
of the continuum $\theta$.  The diffusion equation in all
dimensions and the $d^n x/dt^n=\eta(t)$ processes for $n \ge
3$ are all suitable.  Table \ref{tableoldresults} shows the
continuum results for diffusion in 1-3 dimensions as
reported in \cite{ctl}, with numerical results and also the
singly constrained and IIA results for comparison.  For
$d=1$ the IIA is slightly better, but the correlator
expansion is more accurate for $d=2$,$3$.  Furthermore, we
obtain estimates of the errors and, by going to higher order
we may improve our results.  Table \ref{tablednxresults}
shows the continuum results for the $d^n x/dt^n=\eta(t)$
processes with $3 \le n \le 10$.  Figure
\ref{fdnxplot3to10ANDANDfdnxplotvs1overi} shows how
$\theta_n$ varies with $n$.  Notice in particular that
$\theta_n -\theta_{\infty} \propto 1/n$ for $n>20$, 
and that $\theta_{\infty}$ is identical to that of 2-d
diffusion (since the correlators are identical).

\begin{center}
\begin{tabular}{||l||l|l|l|l||} \hline
 &  Pad\'e1CR &  Pad\'e2CR &  numerical &  IIA \\  \hline 
$\ddot  x$ & 0.2506(5)  & \,\,\,\dots& 1/4 (exact) & 0.2647 \\ 
1-d diff   & 0.119(1)   & 0.1201(3)  & 0.12050(5)  & 0.1203 \\  
2-d diff   & 0.187(1)   & 0.1875(1)  & 0.1875(1)   & 0.1862 \\ 
3-d diff   & 0.24(3)    & 0.237(1)   & 0.2382(1)   & 0.2358 \\ \hline
\end{tabular} 
\end{center}
\label{tableoldresults}
\noindent Table \ref{tableoldresults}.  Results for the
continuum persistence exponent $\theta$ for the random
acceleration problem and the diffusion equation in 1-3
dimensions.  Pad\'e1CR is the Pad\'ed results with 1
constraint, Pad\'e2CR has 2 constraints.  Numerical
\cite{newmanandloinaz} and IIA results are shown for
comparison.
\medskip

\begin{center}
\begin{tabular}{||l||l|l||} \hline 
 n  &   Pad\'e2CR &    IIA 
\\  \hline  
   3  & 0.22022(3) 
& 0.22283  
\\ 4  & 0.20958(3) 
& 0.21029  
\\ 5  & 0.20413(3) 
& 0.20417  
\\ 6  & 0.20084(3) 
& 0.20054
\\ 7  & 0.19864(3) 
& 0.19813  
\\ 8  & 0.19707(3) 
& 0.19642  
\\ 9  & 0.19589(3) 
& 0.19514  
\\ 10 & 0.19496(3) 
& 0.19414  
\\ \hline
\end{tabular} 
\end{center}
\label{tablednxresults}
\noindent Table \ref{tablednxresults}.  Results for
$\theta_n$ against $n$ for small $n$.  The Pad\'ed
correlator expansion with 2 constraints is shown along with
the Independent Interval Approximation.  The Pad\'ed results
are an average of suitable Pad\'es of order 14 to 10.  Note
that for $n=2$ (the random acceleration problem), the IIA
gives $\theta=0.2647$ whilst the analytical result is $1/4$.
For $n \to \infty$ (the diffusion equation), the Pad\'ed
result with 2 constraints is $0.1875(1)$, the numerical
result is $0.1875(1)$ and the IIA result is $0.1862$.
\medskip

  \widetext

And so, by the use of these two constraints we have been
able to extend a series expansion about $\Delta\!T = \infty$
all the way to the $\Delta\!T \to 0$ limit.  The ability to
do this does however depend on the correlator.  First, if
the process is `rough', i.e. $1-C(T) \propto T^\beta +
\dots$ with $0 < \beta < 2$, so that the probability distribution 
of the time $T$ between two zero-crossings behaves as  
$P_1(T) \propto T^\alpha + \dots$, with $\alpha < 0$, 
then we have shown \cite{ebm} that $\theta(0) -\theta(\Delta\!T) 
\sim \Delta\!T^{1+\alpha}$ for small $T$.  For the
random walk, for example, $\alpha= - 1/2$ and so we get a square
root cusp in $\theta_D(\Delta\!T)$ for $\Delta\!T \to 0$ 
which the series expansion about $\Delta\!T= \infty$ cannot 
reproduce.  Consider processes such as

\begin{equation}  
{\partial h \over \partial t} = -(-\nabla^2)^{z/2} h +\eta(t)
\label{lingeqn}  
\end{equation} 
describing linear interface growth, where $\eta({\bf x},t)$ is 
delta correlated in space and time. The normalized autocorrelation 
function of $h({x},t)$ has, for $d<z$, the form  
\begin{equation}  
C(T) = \cosh(T/2)^{2 \beta} -|\sinh(T/2)|^{2 \beta} = 1-
|T/2|^{2 \beta} + \dots
\label{lingcorr}  
\end{equation} 
where $T=\ln t$ as usual, $\beta =(1/2) (1-d/z)$ and $d$ is the 
spatial dimension \cite{persistenceoffluctuatinginterfaces}.  
Since $\beta < 1/2$ this process will always be `rough', so 
extrapolation of the series to the continuum limit is not possible. 
Secondly, the correlator may not be easily expandable in
some suitable variable such as the $e^{-\Delta\!T/2}$ used
earlier.  Another process, fractional Brownian motion, defined 
as a Gaussian process $x(t)$ with stationary increments 
$\left<[x(t) - x(t')]^2\right> \propto |t-t'|^{2\beta}$, 
has normalized correlator \cite{fractionalbrownianmotion}
\begin{equation}  
C(T) = \cosh(\beta T)-{1 \over 2} \left|2 \sinh \left({T
\over 2} \right) \right|^{2 \beta}, 
\label{dummy17}  
\end{equation}
where $T=\ln t$ as usual. For general $\beta$ there are  two 
incommensurate variables $e^{-\beta \Delta\!T}$ and $e^{-\Delta\!T/2}$, 
making it difficult to construct a controlled expansion. 

Finally, when applying the 2 constraints to the 3-dimensional diffusion
problem, we have been unable to numerically solve the 44 simultaneous 
nonlinear equations required to construct the expansion to $O(a^{42})$. 
Thus we have only gone up to $a^{29}$ in this problem. This is not
however an insuperable difficulty.

Note however that, even when we cannot get continuum results, for
$\Delta\!T$ large the expansion will always work, as just substituting 
in the raw correlator is good enough.

Having introduced the correlator expansion for the
calculation of persistence exponents, in the following
sections we will extend it to calculate properties of the
occupation-time and crossing-number distributions.

\section{Occupation-time distribution} 
\label{Residenz-time distribution}

The occupation-time distribution, considered for the
continuous case in
\cite{dornic,newman,dharandmajumdardxresidenz},
is the probability distribution $R(\tau(T))$, where
\begin{equation} 
\tau(T)={1 \over T} \int_0^T dT' \Theta(X(T'))
\label{res10}
\end{equation}
and $\Theta(X)$ is the Heaviside step function.  For a
symmetric distribution of zero mean, $R_T(\tau)$ is
symmetric about $\tau =1/2$. Then $R_T(0)$ and
$R_T(1)$ give the persistence probability $P(T)$ introduced
earlier.  The discrete-sampling equivalent, $R_{n,s}$, is
the probability that $X(T)$ has been found to be positive at
exactly $s$ out of the $n$ samplings.  Thus,

\begin{equation} 
s(n)/n = r(n) = {1 \over n}\sum_{i=1}^n \Theta [X(i \Delta \! T)]\ . 
\label{res20} 
\end{equation}
Writing $\Theta [X(i \Delta \! T)]=(1+\sigma_i)/2$, where
$\sigma_i \equiv {\rm sign}[X(i \Delta \! T)]$,  we get

\begin{equation} 
\left< r(n) \right> = {1 \over 2}
\label{res30} 
\end{equation}
and

\begin{equation} 
\left< r(n)^2 \right> = {1 \over 4} + {1 \over 2 \pi n^2}
\sum_{i=1}^n \sum_{j=1}^n \sin^{-1} [C(|i-j| \Delta\!T)]
\label{res40} 
\end{equation}
where we have used the result that $\left< \sigma_i \sigma_j
\right> = (2/\pi) {\sin^{-1}}[C(|i-j| \Delta\!T)]$. If we
choose as before to work to a given order in the correlator,
we need only evaluate the sum up to that order.  Taking the
large $n$ limit, we can change the sum to:

\begin{equation} 
\left< r(n)^2 \right> = {1 \over 4} + {1 \over 4n} 
+ {1 \over \pi n}
\sum_{k=1}^{o} \sin^{-1}[C(k \Delta\!T)]
\label{res50} 
\end{equation}
where $o$ is the order to which we wish to work.  It has
been pointed out \cite{mandbresidenzetc} that for $n$ large,
the widely separated (in time) parts of the time series
become uncorrelated and, following the central limit
theorem, $R_{n,s}$ is Gaussian for $s$ close to $1/2$ with
standard deviation given by equation (\ref{res50}).  We will
use this as a check of our final result for $R_{n,s}$.

To find $R_{n,s}$ we must sum over all `paths' involving $s$
positive samplings (and $n-s$ negative ones). So the probability, 
$R_{n,s}$ to find $s$ positive values from $n$ samplings is 
\begin{equation} 
R_{n,s} = \left<\delta_{2s-n,\sum_i \sigma_i}\right> 
= \left< {1 \over 2^n} \sum_{\{\epsilon_i=\pm 1\}}
\delta_{2s-n,\sum_i \epsilon_i} \prod_{i=1}^n (1+\epsilon_i
\sigma_i) \right>
\label{res60} 
\end{equation}
where $\delta_{\alpha,\beta}$ is the Kronecker delta
function which we choose to write in analytic form as a
Cauchy integral
\begin{equation}
\delta_{\alpha,\beta}={1 \over 2 \pi i} \oint {dz \over
z^{\alpha-\beta+1}},
\label{res70} 
\end{equation}
where the integration contour encircles the origin. 
Substituting this into eqn. (\ref{res60}) gives,
\begin{equation} 
R_{n,s} = {1 \over 2 \pi i} \oint {dz \over z^{2s-n+1}}
\left< {1 \over 2^n} \sum_{\{\epsilon_i=\pm 1\}} \prod_{i=1}^n
z^{\epsilon_i} (1+\epsilon_i \sigma_i) \right>.
\label{res80} 
\end{equation}
Summing over the $\epsilon_i$s gives:

\begin{equation} 
R_{n,s} = {1 \over 2 \pi i} \oint {dz \over z^{2s-n+1}}
\left({1+z^2 \over z}\right)^n \left< {1 \over 2^n}
\prod_{i=1}^n \left(1+ {z^2 -1 \over z^2 +1} \, \sigma_i
\right) \right> .
\label{res90} 
\end{equation}
The term which is averaged over is identical to that of the
normal persistence calculation apart from the factor
$(z^2-1)/(z^2+1)$ associated with each $\sigma_i$.  Making
this minor change to the previous calculation of persistence
gives us a term $\tilde{\Upsilon}^n(z)$ where before we had
$\rho^n$, and so $\tilde{\Upsilon}(\infty)=\rho$.  Replacing 
$s$ by $rn$ where $0\le r \le 1$ and anticipating that $R_{n,s}
\sim [\rho(r)]^n$ for $n$ large gives us an expression for
$\rho(r)$ which we can evaluate by steepest descents:

\begin{equation} 
[\rho(r)]^n = {1 \over 2 \pi i} \oint {dt \over 2 t} \, {\rm
exp}[n({\rm ln}(1+t)-r{\rm ln}t +{\rm ln}\Upsilon(t))]
\label{res100} 
\end{equation}
where we have replaced $z^2$ by $t$, and 
$\Upsilon(t) = \tilde{\Upsilon}(\sqrt{t})$. As a simple check, 
at zeroth order $\Upsilon$ is $1/2$ and the method of steepest
descents gives a saddle-point value $t_s^{(0)} = r/(1-r)$, and 
\begin{equation} 
R_{n,s} \sim  {1 \over 2^n} \exp\{-n[r\ln r + (1-r)\ln (1-r)]\}.
\label{res110} 
\end{equation}
This is the same as the combinatorial result,

\begin{equation} 
R_{n,s} = {1 \over 2^n} {n \choose rn}
\label{res120} 
\end{equation}
when expanded to leading order for large $n$ using Stirling's
formula.  Note that there is hence also a $\sqrt{n}$ term in
$R_{n,s}$ which we ignore relative to the exponential for $n
\to \infty$.  

We use the method of steepest descents in the following way. 
Having found the position $t_s^{(0)}$ of the saddle point to zeroth 
order, we substitute it into the right-hand side of the general 
saddle-point equation 
\begin{equation}  
t_s = {r \over {1-r}} -{t_s(1+t_s) \over 1-r} \Upsilon'(t_s),
\label{dummy18}  
\end{equation}
where $\Upsilon'(t) \equiv d\Upsilon/dt$, and thus find $t_s$ to first 
order, and so on recursively up to 10th order.  Substituting $t_s$ 
into the exponent of equation (\ref{res100}) gives an analytic expression 
for $R_{n,s}$ in the large $n$ limit and hence $\rho(r)$. Just as in 
the expression for persistence, the expression for $\rho(r)$ is 
rather long and it was only possible to find $\rho(r)$ analytically 
to 10th order.

As stated in the previous section, for $r$ close to
$<r>=1/2$, $\rho(r)$ approximates to a gaussian distribution,

\begin{equation}  
\rho(r) \propto \exp \left[ -{1 \over 2 n} {(r-\left< r \right>)^2 
\over \left<r^2\right>-\left<r\right>^2} \right] .
\label{dummy19}  
\end{equation}
Thus one expects that the quantity 
$\lim_{r \to \left<r\right>} (r - \left<r\right>)^2/[-2 n \ln \rho(r)]$ 
should equal the variance of $r$ calculated previously, and indeed 
these two quantities agree term by term to 10th order,
providing a useful cross-check. 

We apply our result to the random walk, random acceleration,
and diffusion from random initial conditions in 1-3
dimensions. We recall that $R_{n,s}$ is the  probability for 
$n$ measurements of $X$ to return $s$ positive values. We have shown 
that for $n \to \infty$, $s \to \infty$ with $r=s/n$ fixed it has the 
form $R_{n,s} \sim [\rho(r)]^n$, which can be written in the alternative 
form $R_{n,s} \sim \exp[-\theta_D(r)T]$, where $T=n\Delta T$ as usual 
and $\theta_D = -\ln\rho(r)/\Delta T$. 

\begin{figure} 
\narrowtext \centerline{\epsfxsize\columnwidth
\epsfbox{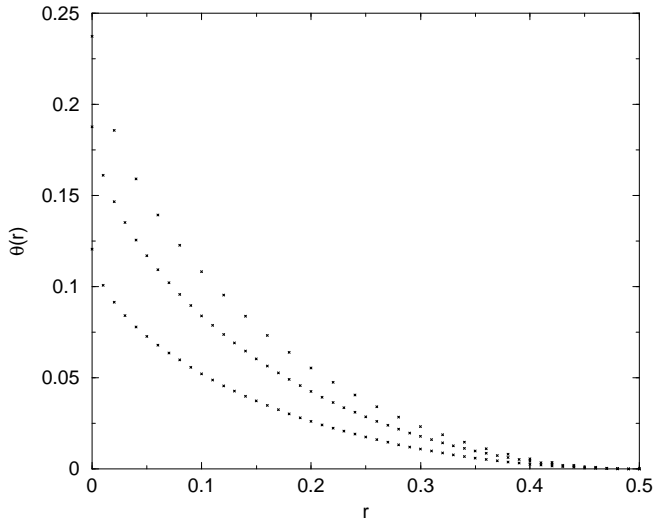}}
\caption{Plot of the continuum large deviation function
$\theta(r)$ against $r$ for the diffusion equation in 1, 2
and 3 dimensions (bottom to top respectively).  $\theta(r)$
is symmetric about $r=1/2$.  The results were obtained using
Pad\'es with 2 constraints, an average being taken of
suitable Pad\'es of order 10 to 7 for 1 and 2 dimensions and
of order 7 to 6 for 3 dimensions.}
\label{fresdiff123continuumthetavsr}

\end{figure}

\begin{figure} 
\narrowtext \centerline{\epsfxsize\columnwidth
\epsfbox{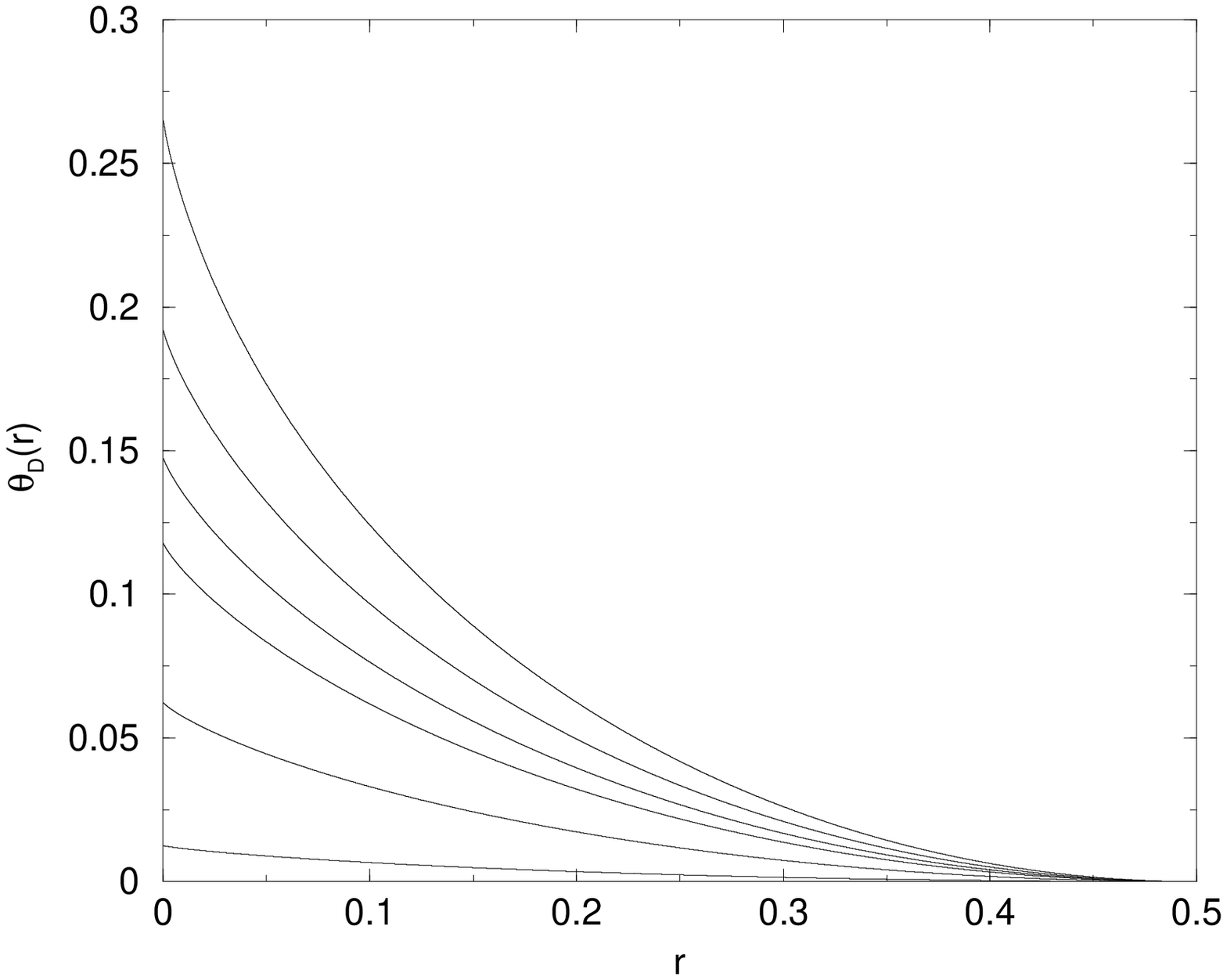}}
\caption{Plot of $\theta_D(r)$ against $r$ for the random walk
with ${\rm exp}(-\Delta\!T/2)=1/2$, $1/4$, $1/8$, $1/16$,
$1/256$ and $1/2^{40}$ (top to bottom respectively).  The
curves were plotted from the raw series in ${\rm
exp}(-\Delta\!T/2)$ to 10th order.  Note that $r=0,1$
corresponds to ordinary discrete persistence and that the
curves are symmetric about $r=1/2$.  }
\label{fresdxdiscrete}

\end{figure}

\begin{figure} 
\narrowtext \centerline{\epsfxsize\columnwidth
\epsfbox{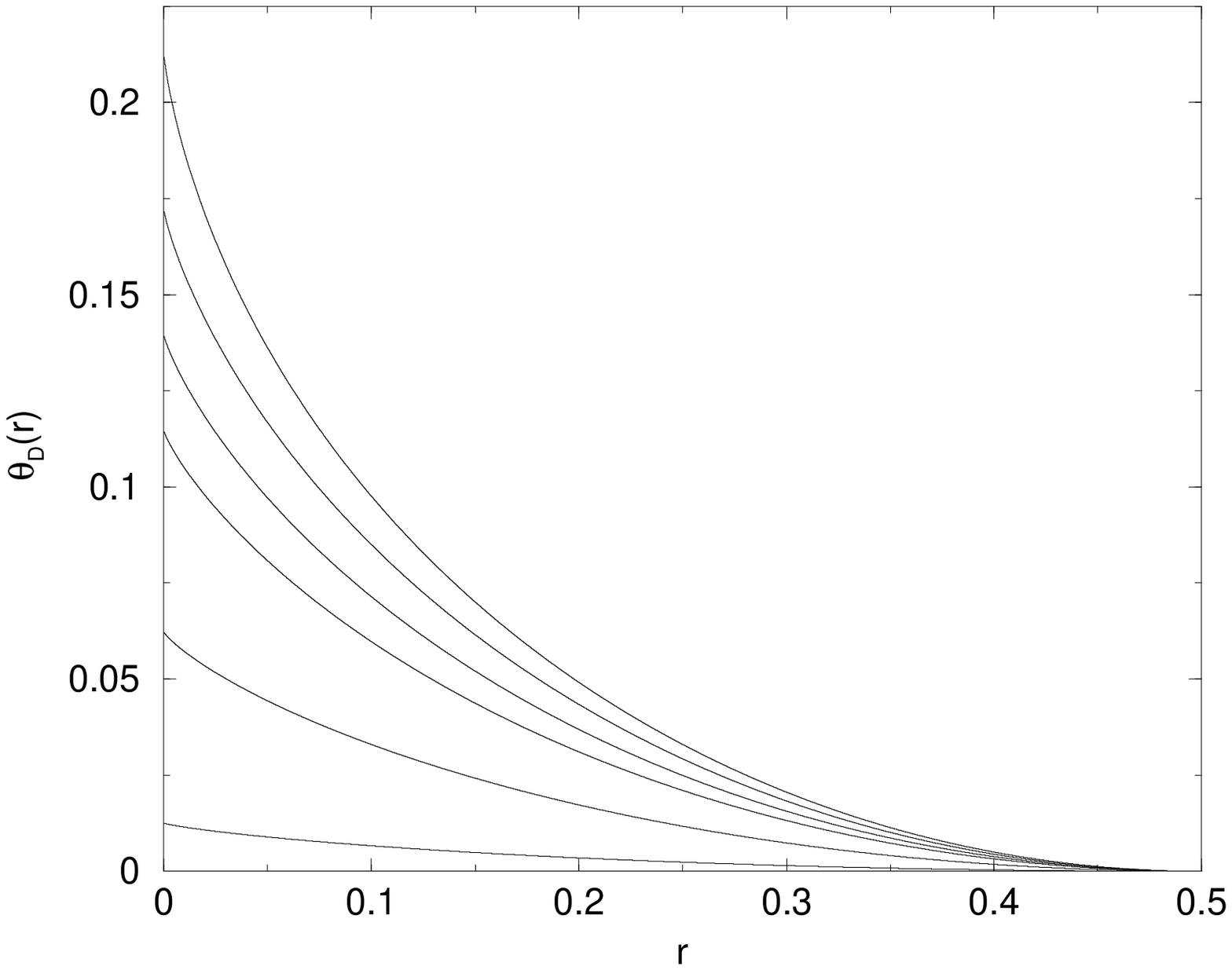}}
\caption{Plot of $\theta_D(r)$ against $r$ for the random
acceleration with ${\rm exp}(-\Delta\!T/2)=1/2$, $1/4$,
$1/8$, $1/16$, $1/256$ and $1/2^{40}$ (top to bottom
respectively).  The curves were plotted from the raw series
in ${\rm exp}(-\Delta\!T/2)$ to 10th order.  Note that
$r=0,1$ corresponds to ordinary discrete persistence and
that the curves are symmetric about $r=1/2$.  }
\label{fresddxdiscrete}

\end{figure}

\begin{figure} 
\narrowtext \centerline{\epsfxsize\columnwidth
\epsfbox{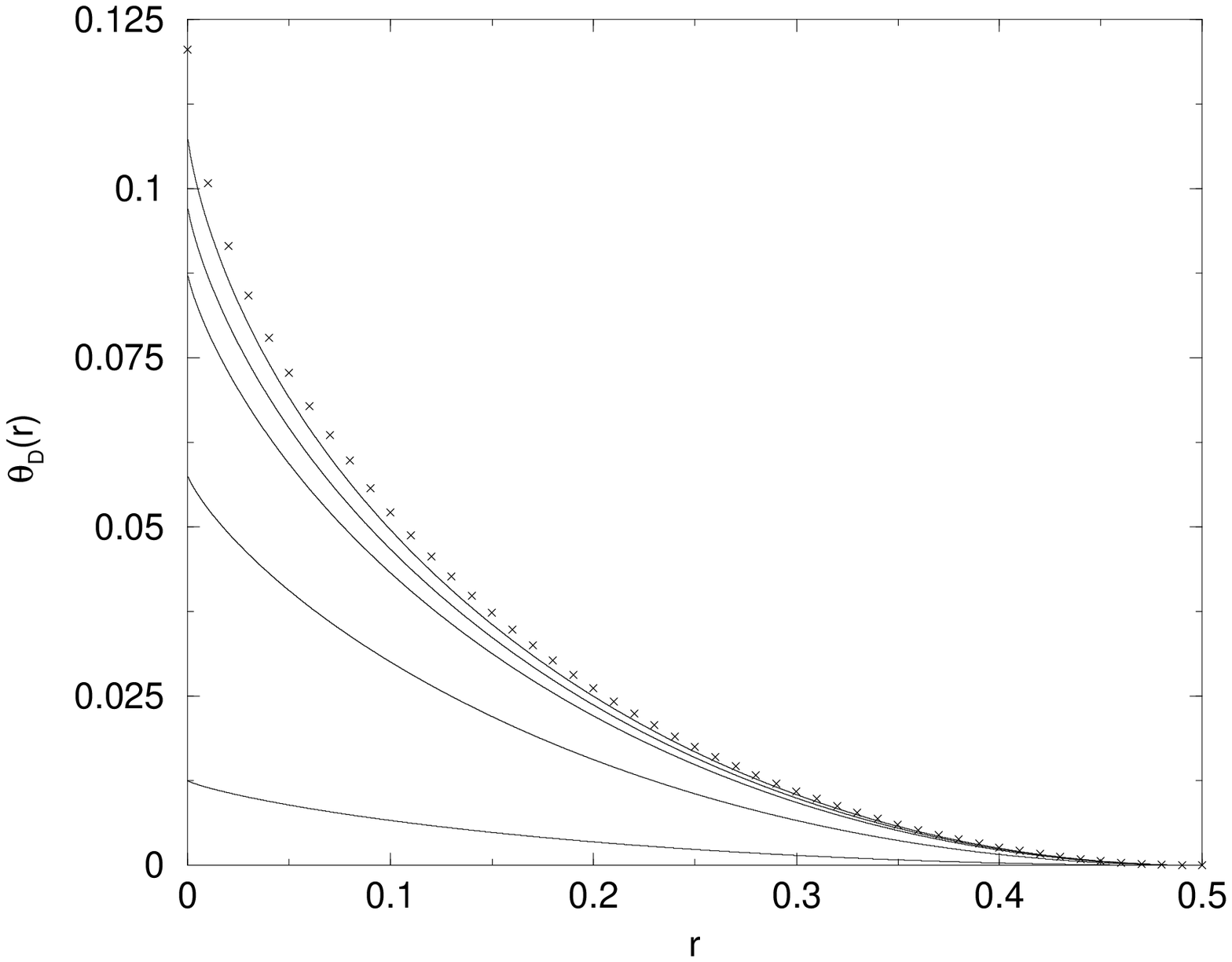}}
\caption{Plot of $\theta_D(r)$ against $r$ for diffusion in 1
dimension with ${\rm exp}(-\Delta\!T/2)=1/4$, $1/8$, $1/16$,
$1/256$ and $1/2^{40}$ (top to bottom respectively).  The
curves were plotted from the raw series in ${\rm
exp}(-\Delta\!T/2)$ to 10th order.  The ${\rm
exp}(-\Delta\!T/2)=1/2$ curve is not shown as the raw series
had not converged at 10th order.  Also shown are the results
for the continuum limit from figure
\ref{fresdiff123continuumthetavsr}.  Note that $r=0,1$
corresponds to ordinary discrete persistence and that the
curves are symmetric about $r=1/2$.}
\label{fresdiff1ddiscrete}

\end{figure}

\begin{figure} 
\narrowtext \centerline{\epsfxsize\columnwidth
\epsfbox{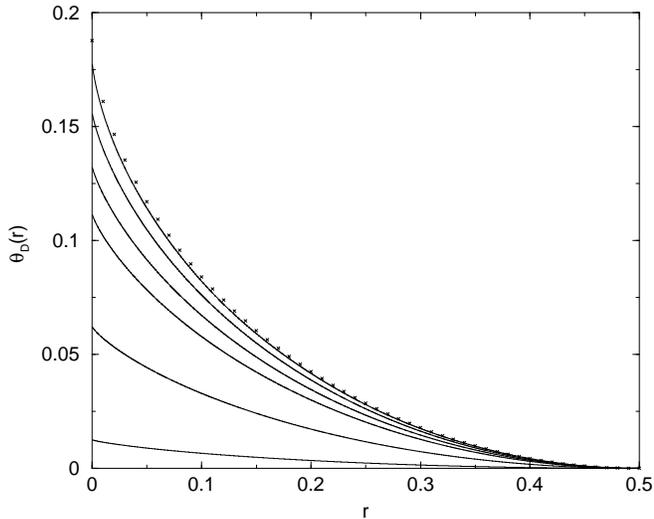}}
\caption{Plot of $\theta_D(r)$ against $r$ for diffusion in 2
dimension with ${\rm exp}(-\Delta\!T/2)=1/2$, $1/4$, $1/8$,
$1/16$, $1/256$ and $1/2^{40}$ (top to bottom respectively).
The curves were plotted from the raw series in ${\rm
exp}(-\Delta\!T/2)$ to 10th order.  Also shown are the results
for the continuum limit from figure
\ref{fresdiff123continuumthetavsr}.  Note that $r=0,1$
corresponds to ordinary discrete persistence and that the
curves are symmetric about $r=1/2$.  }
\label{fresdiff2ddiscrete}

\end{figure}

\begin{figure} 
\narrowtext \centerline{\epsfxsize\columnwidth
\epsfbox{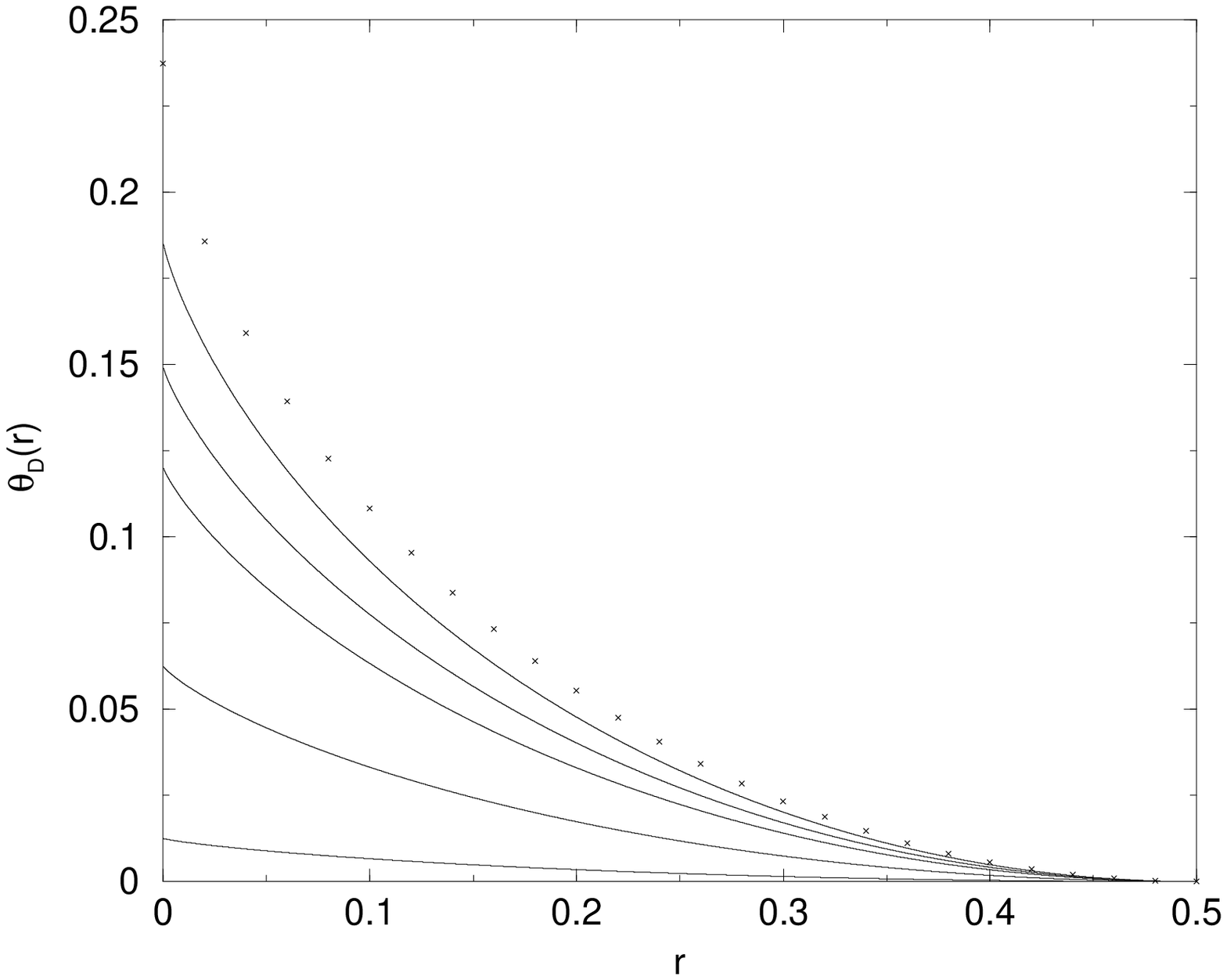}}
\caption{Plot of $\theta_D(r)$ against $r$ for diffusion in 3
dimension with ${\rm exp}(-\Delta\!T/2)=1/4$, $1/8$, $1/16$,
$1/256$ and $1/2^{40}$ (top to bottom respectively).  The
curves were plotted from the raw series in ${\rm exp}(-3
\Delta\!T/2)$ to 10th order.  The ${\rm
exp}(-\Delta\!T/2)=1/2$ curve is not shown as the raw series
had not converged at 10th order.  Also shown are the results for
the continuum limit from figure
\ref{fresdiff123continuumthetavsr}.  Note that $r=0,1$
corresponds to ordinary discrete persistence and that the
curves are symmetric about $r=1/2$.  }
\label{fresdiff3ddiscrete}

\end{figure}

Plots of $\theta_D(r)$ against $r$ for various
values of $\Delta\!T$ are shown in figures
\ref{fresdxdiscrete}, \ref{fresddxdiscrete},
\ref{fresdiff1ddiscrete}, \ref{fresdiff2ddiscrete}
and \ref{fresdiff3ddiscrete}.
For the diffusion equation we are able to Pad\'e the series
and apply 2 constraints, giving us good estimates for the
continuum $\theta(r)$, i.e.\ the limiting value of $\theta_D(r)$ as 
$\Delta T \to 0$.  Plots of $\theta(r)$ are also shown in
figure \ref{fresdiff123continuumthetavsr}.  The second
constraint, that $d \theta(r)/ d \Delta\!T |_{\Delta\!T=0}
=0$ for sufficiently smooth processes (including diffusion), 
comes from a similar argument to that given earlier \cite{ebm} 
for standard persistence: as $\Delta\!T$ is increased from zero, 
the first correction to $\theta$ comes from the contribution 
of a path that is the same as a
contributing path in the continuum, apart from one
undetected double crossing which (to lowest order in
$\Delta\!T$) gives a correction to $\theta$ of order
$\Delta\!T^2$ and thus $d \theta(r)/ d \Delta\!T
|_{\Delta\!T=0} =0$.

The function $\theta(r)$ is the large-deviation function for 
the occupation-time distribution. Close to $r = \left<r\right> = 1/2$, 
it is quadratic in $r-\left<r\right>$. The probability distribution 
$P_r(r)$ of $r$ is given by $P_r(r) \propto [\rho(r)]^n = 
\exp[-(1/2)(r -\left<r\right>)^2/(\left<r^2\right>-\left<r\right>^2)$ 
for $r$ near $\left<r\right>$. This means that the typical fluctuations in $r$ 
around the mean are of order $n^{-1/2}$ for large $n$ since the variance 
is proportional to $1/n$. The full function  
$\theta(r)$ is required to determine the probability of large deviations 
from the mean, where the fluctuations are non-Gaussian.

\begin{figure} 
\narrowtext \centerline{\epsfxsize\columnwidth
\epsfbox{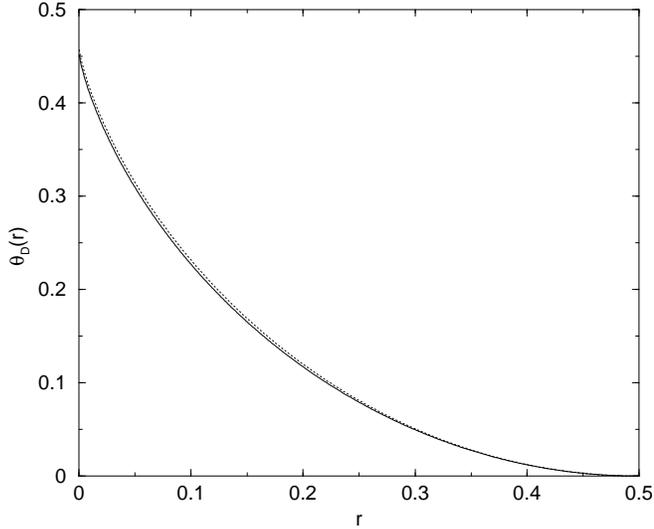}}
\caption{Plot of $\theta_D(r)$ against $r$ for the intrinsically 
discrete process $\psi_i=(\phi_i+\phi_{i-1})/\sqrt{2}$ where the 
$\phi_i$s are independent identically distributed symmetric random 
variables.  The curves show the raw correlator result to 10th order 
(dotted line) and the exact result [18] (solid line).  The curves 
differ by a maximum of 0.00575.  Note that $r=0,1$ corresponds to ordinary 
discrete persistence and that the curves are symmetric about $r=1/2$.}
\label{foccupationtime_May2003}

\end{figure}

We end this section by noting that the full large deviation function 
$\theta_D(r)$ associated with the occupation-time distribution was computed 
analytically\cite{md} for the intrinsically discrete process
\begin{equation}
\psi_i = \cos(\omega) \phi_i +\sin(\omega) \phi_{i-1}\ ,
\label{satyasprocess}
\end{equation}
where the $\phi_i$ are independently distributed gaussian random
variables. This process appears as a limiting case of the diffusion equation on
a hierarchical lattice\cite{mdhar} and also appears in the one dimensional
Ising spin glass problem\cite{md,dg}. Exact results were
obtained for the case $\omega = \pi/4$.
Interestingly, these results turn out to be independent of the
distribution of $\phi_i$ provided that it is symmetric. We have also
obtained the the large deviation function for $\omega=\pi/4$ by the 
correlator method. The comparison with the exact results is shown in 
figure \ref{foccupationtime_May2003}.

  \widetext

  \widetext

  \widetext

  \widetext

  \widetext

  \widetext

  \widetext

\section{Occupation-Time Partial Survival} 
\label{Residenz Partial Survival}

Here we consider the discrete occupation-time partial-survival
probability, $R_n(p)$. Let us suppose that the process `dies' 
with probability $1-p$ whenever $X$ is sampled to be positive.  
Then $R_n(p)$ is defined to be the probability of the process 
surviving $n$ samplings if the variable $X_i$ survives being sampled
as positive with probability $p$.  Samplings as negative are
always survived. Thus,

\begin{equation} 
R_n(p) = \sum_{s=0}^n p^s R_{n,s}
\label{res130} 
\end{equation}
and so $R_n(p)$ is also the generating function for
$R_{n,s}$ since

\begin{equation} 
R_{n,s} = {1 \over s!} {d^s \over dp^s}|_{p=0} R_n(p)
\label{res140} 
\end{equation}
or alternatively

\begin{equation} 
R_{n,s} = \oint {dp \over p^{s+1}} R_n(p)
\label{res145} 
\end{equation}
Also, writing $p^s$ as ${\rm exp} ( s {\rm ln} p )$ and
expanding the exponentials gives:

\begin{equation} 
{\rm ln} \rho^r(p) = \sum_{j=1}^\infty {{\rm ln} p^j \over
j!} \left<s^j\right>_c
\label{res150} 
\end{equation}
where $< s^j >_c$ is the $j$th cumulant of the occupation time, $s$, 
and we have used $R_n(p) = [\rho(p)]^n$, which,
as for persistence (but unlike $R_{n,s}$) is true for
any $n$ provided that $n$ is larger than the largest diagram
involved in the evaluation of $R_n(p)$.  Thus calculating 
$R_n(p)$ gives us another method for finding the moments 
of the number of crossings and also $R_{n,s}$ although the
evaluation of $R_{n,s}$ by the contour integration is
entirely equivalent to the previous method and
differentiating $R_n(p)$ $s$ ($=r n$) times becomes unfeasible 
for large $n$. 

\begin{figure} 
\narrowtext \centerline{\epsfxsize\columnwidth
\epsfbox{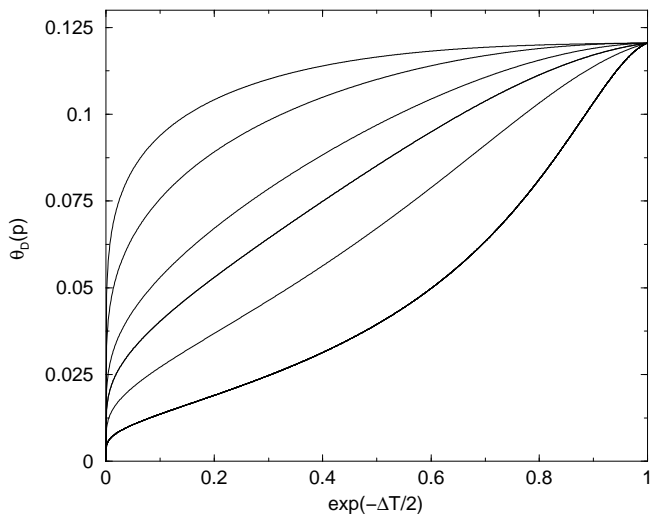}}
\caption{Plot of $\theta_D(p)$ against ${\rm
exp}(-\Delta\!T/2)$ for diffusion in 1 dimension with $p=0$,
$1/4$, $1/2$, $5/8$, $3/4$ and $7/8$ (from the top
respectively).  $\theta_D(p)$ has been constrained to give
the persistence result at the continuum.  The curves are
produced from averages of suitable constrained Pad\'es,
although in practice the various Pade\'s are
indistinguishable.  }
\label{fdiff1dresidenzpartials}

\end{figure}

\begin{figure} 
\narrowtext \centerline{\epsfxsize\columnwidth
\epsfbox{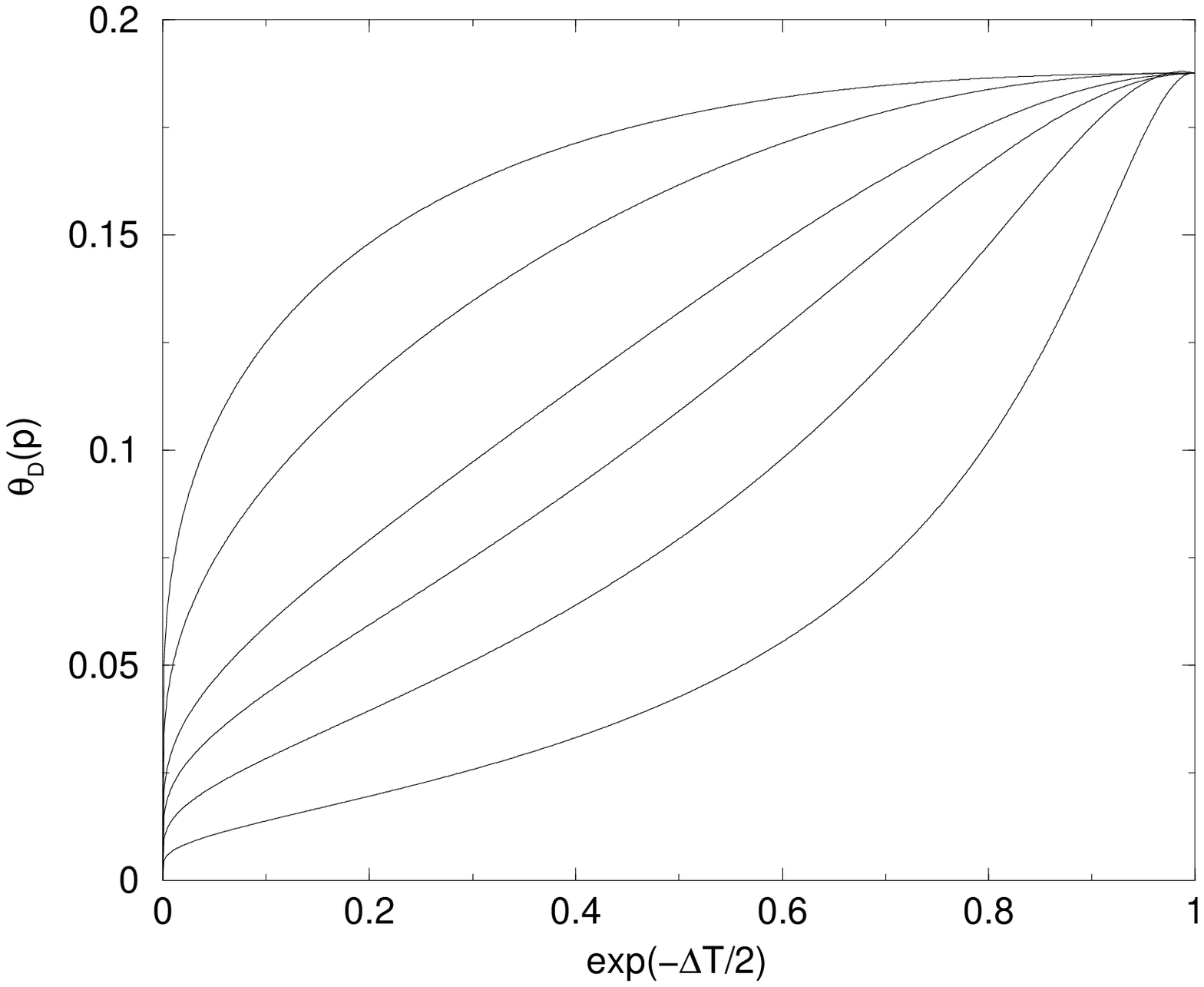}}
\caption{Plot of $\theta_D(p)$ against ${\rm
exp}(-\Delta\!T/2)$ for diffusion in 2 dimensions with
$p=0$, $1/4$, $1/2$, $5/8$, $3/4$ and $7/8$ (from the top
respectively).  $\theta_D(p)$ has been constrained to give
the persistence result at the continuum.  The curves are
produced from averages of suitable constrained Pad\'es,
although in practice the various Pade\'s are
indistinguishable.  }
\label{fdiff2dresidenzpartials}

\end{figure}

\begin{figure} 
\narrowtext \centerline{\epsfxsize\columnwidth
\epsfbox{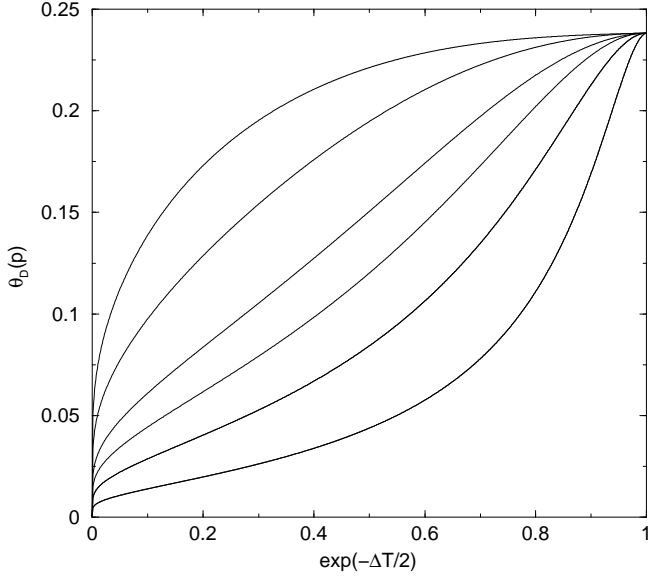}}
\caption{Plot of $\theta_D(p)$ against ${\rm
exp}(-\Delta\!T/2)$ for diffusion in 3 dimensions with
$p=0$, $1/4$, $1/2$, $5/8$, $3/4$ and $7/8$ (from the top
respectively).  $\theta_D(p)$ has been constrained to give
the persistence result at the continuum.  The curves are
produced from averages of suitable constrained Pad\'es,
although in practice the various Pade\'s are
indistinguishable.  }
\label{fdiff3dresidenzpartials}

\end{figure}

$R_n(p)$ is found in a similar way to before, by summing over
all possible `paths',

\begin{equation} 
R_n(p) = \left< {1 \over 2^n} \sum_{\epsilon_i=\pm 1}
\prod_{i=1}^n \left \{\begin{array}{rl} p(1+\sigma_i), &
\qquad \epsilon_i=1 \\ (1-\sigma_i), & \qquad \epsilon_i=-1
\end{array}\right .  \right>,
\label{dummy21} 
\end{equation}
where $\sigma_i = {\rm sgn}(X_i)$ as usual, and the average is 
over the variables $X_i$ ($i=1,\ldots,n$). Thus we get

\begin{equation}  
R_n(p) = \left< {1 \over 2^n} (p+1)^n \prod_{i=1}^n
\left(1+{p-1 \over p+1}\sigma_i \right) \right>
\label{dummy22}  
\end{equation}
which is the same as the calculation for normal persistence
except that we include a factor $(p-1)/(p+1)$ with each
$\sigma_i$ and and overall factor $(p+1)^n$.  Thus we can
find $\rho(p)=\exp[-\theta(p)]$ to order 14.  This is done and results for
diffusion in 1-3 dimensions are shown in figures 
(\ref{fdiff1dresidenzpartials}, \ref{fdiff2dresidenzpartials}, and 
\ref{fdiff3dresidenzpartials}).
Note that, for $\Delta\!T \to 0$, a positive excursion by
the underlying continuous process will survive with zero
probability since the number of samplings $\to \infty$.
Thus $\theta(p)|_{\Delta\!T=0}$ is just the continuum
persistence exponent.  It is therefore possible to improve
$\theta(p)$ for $\Delta\!T$ small by applying
$\theta(p)|_{\Delta\!T=0} =\theta$ as a constraint, in addition
to the standard constraint $\rho(p)|_{\Delta\!T=0}=1$.

  \widetext

  \widetext

  \widetext

A further check is provided by using $\rho(p)$ to generate
the first two cumulants.  The results agree term by term to
10th order with the method used to calculate the mean
(trivially) and the variance.

\begin{figure}
\narrowtext \centerline{\epsfxsize\columnwidth
\epsfbox{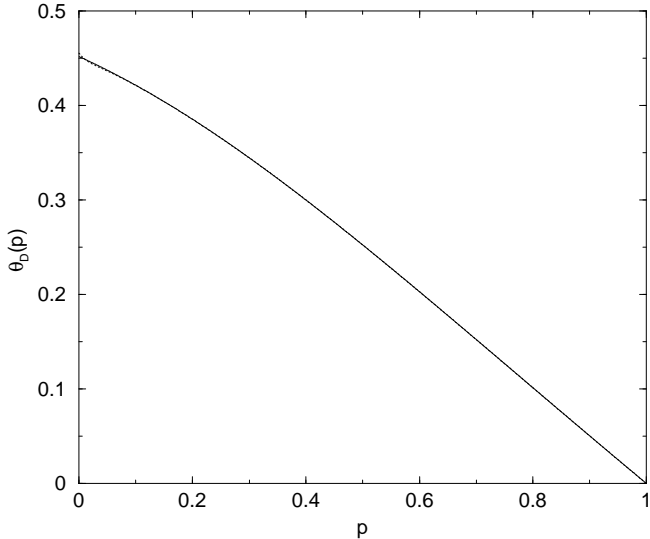}}
\caption{Plot of $\theta_D(p)$ against $p$ for the process 
$\psi_i=(\phi_i+\phi_{i-1})/\sqrt{2}$.  The exact (solid line) and raw 
correlator expansion (dotted line) results are shown, the two curves being 
indistinguishable except near $p=0$ }
\label{foccupationtimePARTIALSURVIVAL_May2003}
\end{figure}

We next compare the results obtained by the correlator method to an
exactly solvable case, namely the discrete process in Eq. (\ref{satyasprocess})
for $\omega=\pi/4$. In this case, an exact expression of the exponent 
$\theta(p)$ is known\cite{md},
\begin{equation}
\theta(p)= {(1-p)\over {2\,{\tan}^{-1}\left({ {1-p}\over {1+p}}\right)}}.
\label{thetapexact}
\end{equation}
A comparison of this exact result with the one obtained by the correlator 
method is shown in figure \ref{foccupationtimePARTIALSURVIVAL_May2003}.

  \widetext

In the last 2 sections we have examined the occupation-time statistics.  
The occupation time depends only on the signs of $X(i
\Delta\!T)$ at each $i$, that is, it is local.  This meant
that we merely had to attach additional factors to each
local $X(i \Delta\!T)$.  In the next sections we will be
studying the number of crossings, so we must consider the
signs of both $X(i \Delta\!T)$ and $X((i+1) \Delta\!T)$.
Thus the problem is not local in the sense used above and we
cannot just attach additional factors to each $X(i
\Delta\!T)$.  The solution, as explained in the next
section, is to attach additional factors to the lines
connecting two $X$s in the diagrammatic notation.

\section{Distribution of Crossings} 
\label{Distribution of Crossings}

\begin{figure} 
\narrowtext \centerline{\epsfxsize\columnwidth
\epsfbox{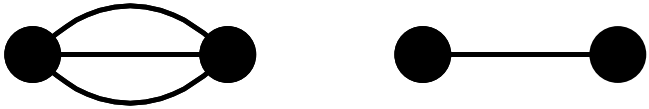}}
\caption{A 4th order diagram.}
\label{fii30}
\end{figure}

We now apply the correlator expansion to calculate the
distribution of crossings of an arbitrary GSP.  
We start from the calculation of the persistence.  The
method is the same up until we assign factors to the
diagrams on the lattice.  We wish to calculate the
probability of $m$ detected crossings in $n$ samplings,
$P_{n,m}$, rather than just the probability of no crossings
which was the persistence calculation.  To do this, we sum
over all the possible ways in which those $m$ crossings
could occur.  Furthermore, we note that if we have a line in
a diagram connecting two vertices, and $s$ crossings occur
between these two vertices, then the factor $C(j \Delta\!T)$
associated with it from the persistence calculation should
also have a factor $(-1)^s$ associated with it.  Consider
the diagram shown in figure \ref{fii30}.

  \widetext

Besides the enumerations done for the persistence
calculation we must consider the following four cases. \\
1: No crossings occur on the sites where the diagram is placed.  
This would occur with probability
\begin{equation} 
{ (n-m-1)(n-m-2) \over (n-1)(n-2) }
\label{dist10} 
\end{equation}
and there are no sign changes. \\
2: There is one crossing between the first and second vertices.  
This occurs with probability
\begin{equation} 
{ m (n-m-1) \over (n-1)(n-2) }
\label{dist20} 
\end{equation}
and there is a factor $(-1)^3$ associated with it. \\ 
3: There is one crossing between the third and fourth vertices.  
The probability of this occurring is as above and there is a
factor $(-1)$ associated with it. \\ 
4: There is one crossing between the first and second vertices and 
one crossing between the third and fourth vertices. This occurs with 
probability
\begin{equation} 
{ m (m-1) \over (n-1)(n-2) }
\label{dist30} 
\end{equation}
and there is a factor $(-1)^3 (-1)$ associated with it.
Hence this diagram has an additional factor
\begin{equation} 
{ (n-m-1)(n-m-2) -2 m (n-m-1) +m (m-1) \over (n-1)(n-2) }
\label{dist40} 
\end{equation}
over and above that for the persistence calculation.  
Furthermore, there is an overall factor
\begin{equation} 
{n-1 \choose m}
\label{dist50} 
\end{equation}
accounting for all the ways in which the $m$ crossings can
occur on the whole lattice.  Note that, although we use the
term `probability', when run over the whole lattice
(multiplied by the factors $(n-2)(n-3)/2!$) each probability 
becomes the exact number of ways in which the corresponding event 
occurs.  Thus by introducing these extra rules when enumerating the 
diagrams on the lattice we are able to calculate $P_{n,m}$, the
probability of exactly $m$ detected crossings occurring in
$n$ samplings.  For $n$ large we expect that, as usual, 
\begin{equation} 
P_{n,m} \sim \rho_m^n
\label{dist60} 
\end{equation}
and so we find $\rho_m$ as
\begin{equation} 
\rho_m = \lim_{n \to \infty} { P_{n+1,m+m/n} \over P_{n,m} }.
\label{dist70} 
\end{equation}

\begin{figure} 
\narrowtext \centerline{\epsfxsize\columnwidth
\epsfbox{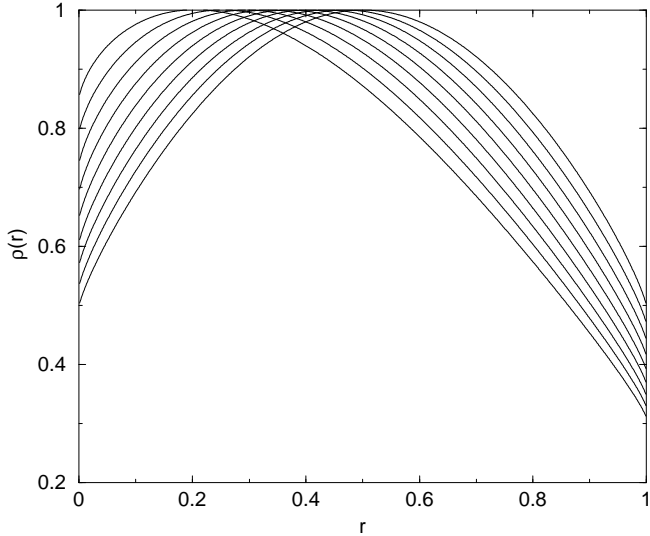}}
\caption{Plot of $\rho(r)$ against $r$ for the random walk
with ${\rm exp}(-\Delta\!T/2)=0$ to $8/10$ in steps of
$1/10$ (top right to bottom right respectively).  Note that
$\rho(r)$ is 1 at the mean value of $r$ given by
$\left<r\right> =1/2 -\sin^{-1}[C(\Delta\!T)]/\pi$.  The
plots are of the raw series.  }

\label{crossings_rhovsr_variousa_dx}
\end{figure}

\begin{figure} 
\narrowtext \centerline{\epsfxsize\columnwidth
\epsfbox{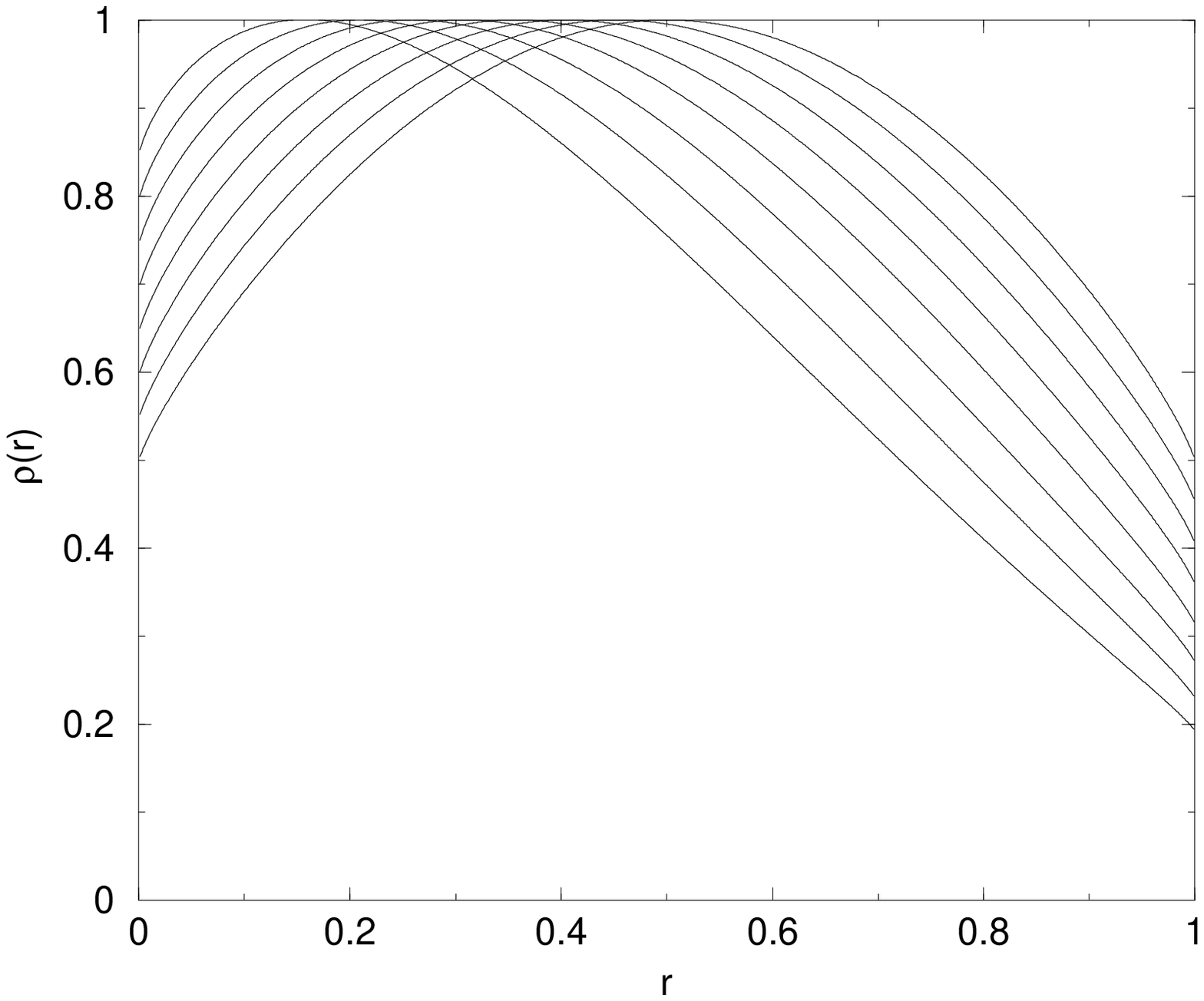}}
\caption{Plot of $\rho(r)$ against $r$ for the random
acceleration process with ${\rm exp}(-\Delta\!T/2)=0$ to
$7/10$ in steps of $1/10$ (top right to bottom right
respectively).  Note that $\rho(r)$ is 1 at the mean value
of $r$ given by $\left<r\right> =1/2
-\sin^{-1}[C(\Delta\!T)]/\pi$.  The plots are of the raw
series.  }

\label{crossings_rhovsr_variousa_ddx}
\end{figure}

\begin{figure} 
\narrowtext \centerline{\epsfxsize\columnwidth
\epsfbox{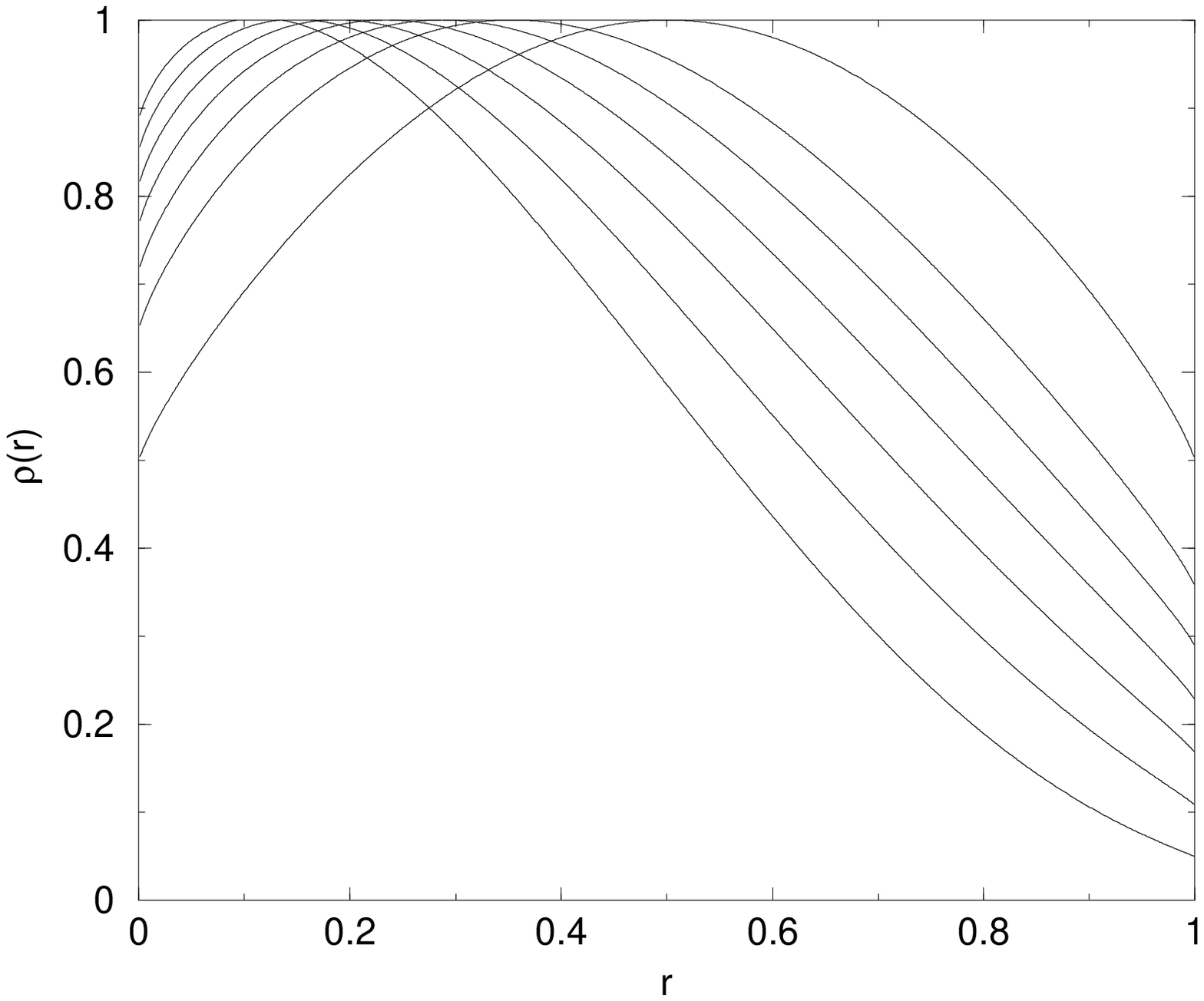}}
\caption{Plot of $\rho(r)$ against $r$ for diffusion in 1
dimension with ${\rm exp}(-\Delta\!T/2)=0$ to $6/10$ in
steps of $1/10$ (top right to bottom right respectively).
Note that $\rho(r)$ is 1 at the mean value of $r$ given by
$\left<r\right> =1/2 -\sin^{-1}[C(\Delta\!T)]/\pi$.  The
plots are of the raw series.  }

\label{crossings_rhovsr_variousa_diff1d}
\end{figure}

\begin{figure} 
\narrowtext           \centerline{\epsfxsize\columnwidth
\epsfbox{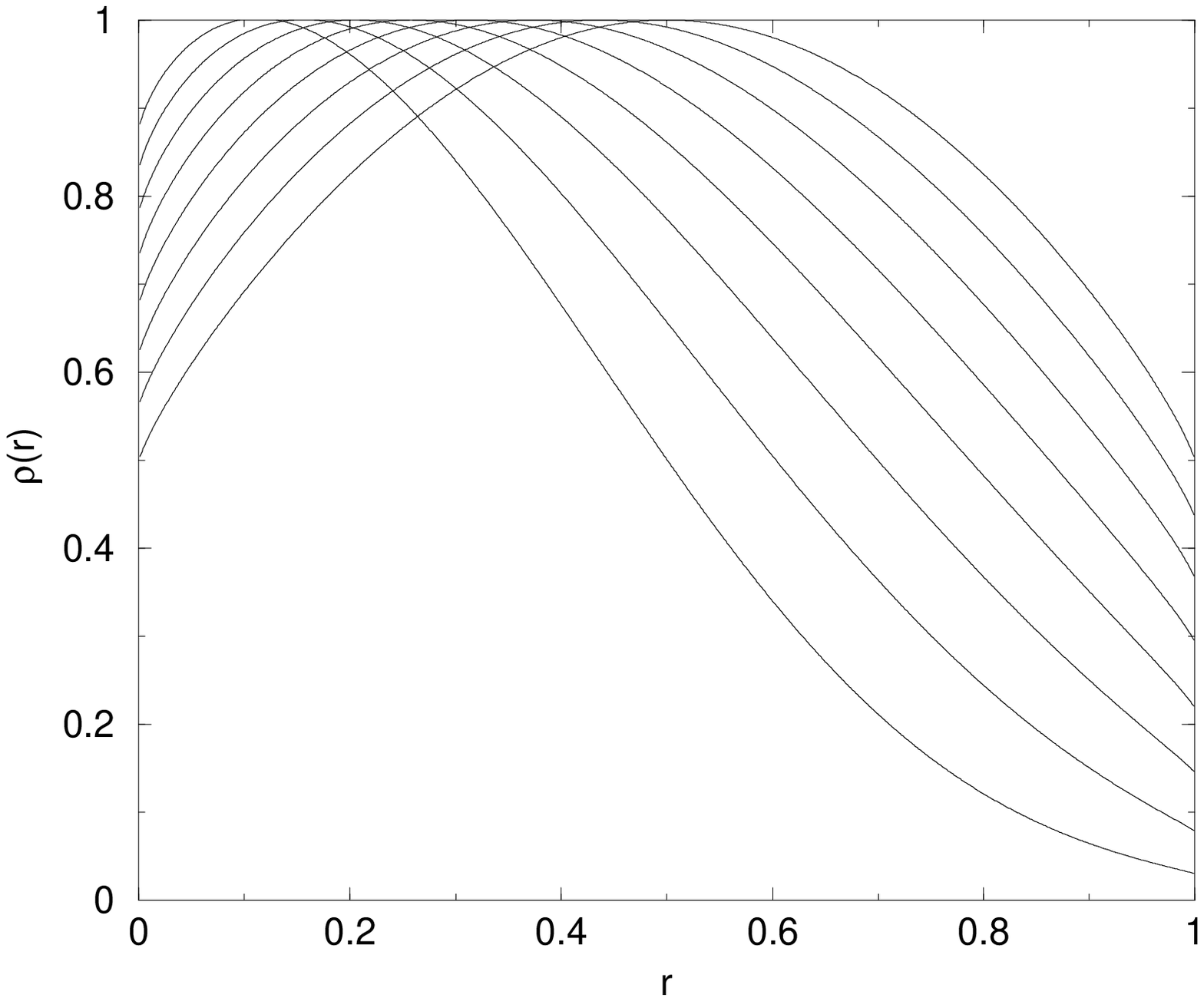}}
\caption{Plot of $\rho(r)$ against $r$ for diffusion in 2
dimensions with ${\rm exp}(-\Delta\!T/2)=0$ to $7/10$ in
steps of $1/10$ (top right to bottom right respectively).
Note that $\rho(r)$ is 1 at the mean value of $r$ given by
$\left<r\right> =1/2 -\sin^{-1}[C(\Delta\!T)]/\pi$.  The
plots are of the raw series.  }

\label{crossings_rhovsr_variousa_diff2d}
\end{figure}

\begin{figure} 
\narrowtext           \centerline{\epsfxsize\columnwidth
\epsfbox{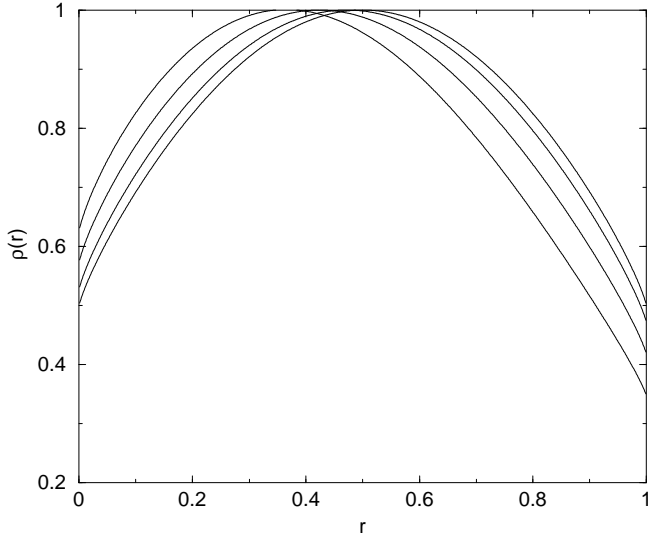}}
\caption{Plot of $\rho(r)$ against $r$ for diffusion in 3
dimensions with ${\rm exp}(-\Delta\!T/2)=0$ to $3/10$ in
steps of $1/10$ (top right to bottom right respectively).
Note that $\rho(r)$ is 1 at the mean value of $r$ given by
$\left<r\right> =1/2 -\sin^{-1}[C(\Delta\!T)]/\pi$.  The
plots are of the raw series.  }

\label{crossings_rhovsr_variousa_diff3d}
\end{figure}

This has been done, although due to the additional factors
it was possible only to go to 10th order due to memory
constraints.  It has been checked that the result agrees
term by term with the normal persistence calculation for
$m=0$ and with alternating persistence for $m=n$.  Figures
\ref{crossings_rhovsr_variousa_dx}, 
\ref{crossings_rhovsr_variousa_ddx}, 
\ref{crossings_rhovsr_variousa_diff1d}, 
\ref{crossings_rhovsr_variousa_diff2d}, and 
\ref{crossings_rhovsr_variousa_diff3d}
show $\rho(r)$ against $r$ where $r n =m$ for various values
of ${\rm exp}(-\Delta\!T/2)$ for the random walk, random
acceleration, and diffusion in 1-3 dimensions.  Notice that,
for $r=\left< r \right>= 1/2 -(1/\pi) \sin^{-1}C(\Delta\!T) $, 
$\rho(r)=1$. Close to this point, $\rho(r)$ approximates to a 
Gaussian distribution,
\begin{equation}  
\rho(r) \propto e^{-{ (r-<r>)^2 \over 2 (<r^2>-<r>^2)}} .
\label{dummy23}  
\end{equation}
where the variance $<r^2>-<r>^2$ agrees term-by-term with the
calculation in the following section. Remember that there 
is a next-to-leading term (pre-exponential factor), $sqrt n$ in $P_{n,m}$, 
as can be seen from considering the $\Delta\!T=\infty$ 
(lowest order) case: 
\begin{equation}  
P_{n,m} = {1 \over 2^n }{n-1 \choose m} \approx {\sqrt{2 \over \pi n}}
\left({1 \over 2} {1 \over (1-r)^{1-r} r^r }\right)^n.
\label{dummy24}  
\end{equation} 

Note that we are considering the number of detected
crossings per sampling (0 or 1). As $\Delta\!T \to 0$ the
fraction of (detected) crossings will go to zero.
Because of this, $\theta(r) \to \infty$ for $\Delta\!T \to
0$ except for the $r=0$ case which is just standard persistence.  
As always, $0 \le \rho(r) \le 1$, and we choose to plot
$\rho(r)$ rather than $\theta(r)$.  We have not applied any
constraints to the series.

  \widetext

  \widetext

  \widetext

  \widetext

  \widetext

\begin{figure} 
\narrowtext           \centerline{\epsfxsize\columnwidth
\epsfbox{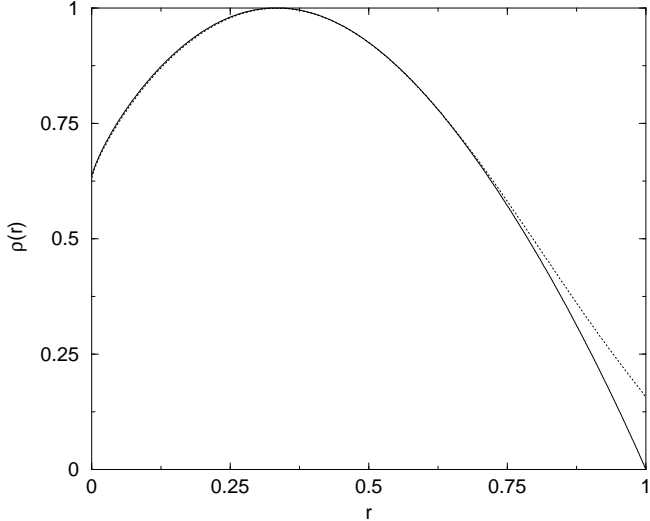  }}
\caption{Plot of $\rho(r)$ against $r$ for the process
$\psi_i=(\phi_i+\phi_{i-1})/\sqrt{2}$ where the $\phi_i$s are
independent identically distributed symmetric random
variables.  The solid line is the numerical evaluation of
the exact result and the dashed line is the result of the
correlator expansion.  The agreement is good until $r
\sim 0.73$, and becomes badly wrong as $r \to 1$ (see text).
}

\label{crossings_rhovsr_variousa_satyasprocess}
\end{figure}

Recently one of us \cite{majumdarpsieqphiiplusphiiminus1}
calculated the distribution of crossings and partial-survival 
probability of the intrinsically discrete process
$\psi_i$ given by Eq. (\ref{satyasprocess}) for the special 
case $\omega=\pi/4$. The correlator of the process is
\begin{equation}  
C(i-j)=\delta_{i,j} + \cos\omega\sin\omega(\delta_{i,j-1}+\delta_{i,j+1}). 
\label{dummy25}  
\end{equation}
Substituting this into the correlator expansion gives the
result shown in figure
\ref{crossings_rhovsr_variousa_satyasprocess} for comparison
with the analytic result.  The agreement is good for $r$
small but for $r \gtrsim 0.73$ the series has not yet
converged by 10th order.  This shows up in the way that
$\rho(r)$ changes as the order is increased from 1 to 10.
For $r$ small there is oscillatory convergence whilst for
$r$ large the convergence is monotonic or, for $r \gtrsim
0.73$, has not occurred.  The fact that convergence does not
occur for $r$ large is presumably because the series is less
good for large numbers of crossings. This also occurs, for
example, for the random acceleration problem where the
alternating persistence ($r=1$) result converges more slowly
than the standard persistence ($r=0$) result. That it fails so 
badly whilst the small $r$ result is acceptable is surprising. 
Nevertheless, by checking whether or not the series converges 
as the order is increased to 10, we can tell whether the result 
is reliable.  For the case $\omega =\pi/12$, for which 
$C(i-j) = \delta_{i,j} + {1 \over 4}(\delta_{i,j-1}+\delta_{i,j+1})$, 
the series has converged for all $r$ although there is no 
analytic result for this case.  
In fact the case studied
is the one for which the correlator takes its largest
possible value.  Also notice that the $\rho(1)=0$ result is
due to the requirement that, in order that $\psi_i$ alternate 
in sign, the magnitude of $\phi_i$ must increase every time
step.  Thus $P_{n,n}$ decays as $2^{-n}/n!$, which is faster than 
a power of $n$, implying $\rho(1)=0$.  For other values of the 
coefficients of $\phi_i$ and $\phi_{i-1}$ this is not the case 
and presumably $\rho(1)$ is non-zero.

  \widetext

In this section we have calculated the distribution of
crossings, $P_{n,m}$, to 10th order in the correlator by
extending the diagrammatic technique.  In the next two
sections we derive the standard deviation of the number of
crossings and then use $P_{n,m}$ to calculate $F_n(p)$, the
partial survival of crossings probability.

\section{The Variance of the Number of Crossings} 
\label{The Variance of the Number of Crossings}

The number of detected crossings in $n$ samplings, $m$, is
(up to an end effect that is negligible for large $n$),

\begin{equation} 
m=r n =\sum_{i=1}^n {1 \over 2} (1-\sigma_i \sigma_{i+1})
\label{varcross10} 
\end{equation}
where $\sigma_i ={\rm sign} [X(i \Delta\!T)]$.  This gives

\begin{equation} 
\left<r\right> = {1 \over 2} -{1 \over \pi}
\sin^{-1}[C(\Delta\!T)]
\label{varcross20} 
\end{equation}
as derived in \cite{barbe}.  One may further attempt to
evaluate the variance of $m$,

\begin{equation} 
\sigma^2/n = \left( \left<m^2\right>-\left<m\right>^2
\right)/n = {1 \over 4 n} \sum_{i=1}^n \sum_{j=1}^n \left<
\sigma_i \sigma_{i+1} \sigma_j \sigma_{j+1} \right> - \left<
\sigma_i \sigma_{i+1} \right> \left<\sigma_j \sigma_{j+1}
\right>\label{varcross30}
\end{equation}
which involves calculating the connected 4-vertex diagrams
in the correlator expansion.  The calculation is essentially
identical to that of section \ref{An introduction to the
correlator expansion} apart from the extra cases of $i=j$
and $i=j \pm 1$ and the result to 14th order may be read
off.  Figures
(\ref{crossings_rhovsr_variousa_dx}
,\ref{crossings_rhovsr_variousa_ddx}
,\ref{crossings_rhovsr_variousa_diff1d}
,\ref{crossings_rhovsr_variousa_diff2d}
,\ref{crossings_rhovsr_variousa_diff3d}
,\ref{crossings_rhovsr_variousa_satyasprocess})
show $\rho(r)$ against $r$ for various processes and values
of $\Delta\!T$.  Close to $r=\left< r \right>$, $\rho(r)$ is
given by

\begin{equation} 
\rho(r) \sim \exp \left[ -{1 \over 2n} {(r-\left< r \right>)^2
\over \left<r^2\right>-\left<r\right>^2} \right]
\label{varcross40} 
\end{equation}
and comparison of $(r-\left< r \right>)^2 /[-2 n \ln
\rho(r)]$ agrees term by term to 14th order in the
correlator with the direct calculation of
$\left<r^2\right>-\left<r\right>^2$, providing a useful
cross-check.

\begin{figure} 
\narrowtext \centerline{\epsfxsize\columnwidth
\epsfbox{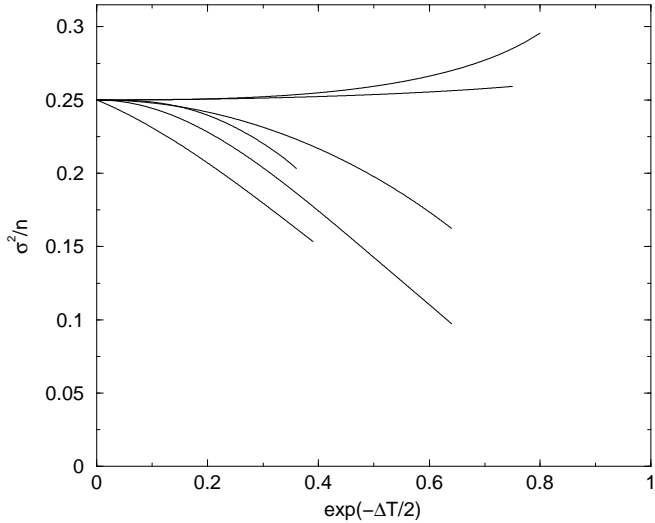  }}
\caption{Plot of $\sigma^2$ against $\exp(-\Delta\!T/2)$.
The curves are, from the top right; the linear growth
equation (\ref{lingeqn}), the random walk, random
acceleration, and diffusion in 3, 2 and 1 dimensions.  
The curves are the raw
series in powers of $\exp(-\Delta\!T/2)$ to 14th order in
the correlator and are plotted only as far as their series have converged.
}
\label{fcrossings_variance}

\end{figure}

The result for $\left<r^2\right>-\left<r\right>^2$ for the
random walk also agrees with that of equation (\ref{var1})
(the matrix method).  The variance for various processes is
shown in figure \ref{fcrossings_variance}.

  \widetext

Thus we have found the variance of the number of crossings for 
an arbitrary process to 14th order in the correlator.  This
involved calculating the 4-vertex diagrams only, and
therefore it is entirely feasible to go to higher orders
since the 4-vertex diagrams are relatively simple.  Notice
also that the variance only contains even orders, as one
would expect from the 4-vertex diagrams.

In the following section we complete our calculations by
finding the partial-survival probability for an arbitrary
GSP, this also being the moment generating function.  The
results will be shown to agree with those of the current
section.

\section{Distribution of Crossings Partial Survival} 
\label{Distribution of Crossings Partial Survival}

As for the specific case of the random walker (Section
\ref{Partial Survival}), we may consider the partial-survival 
probability $F_n(p)$, the probability of surviving
up to the $n$th sampling if each detected crossing is survived with
probability $p$.  As stated in Section \ref{Partial
Survival}, this is also the generating function for
$P_{n,m}$ and the cumulant generating function
\cite{mandbpartialsurvival}:

\begin{equation} 
F_n(p) = \sum_{m=0}^n p^m P_{n,m}
\label{distpart10} 
\end{equation}
and

\begin{equation}  
\ln F(p) = \sum_{j=1}^{\infty} { (\ln p)^j \over j! } \left< r^j \right>_c
\label{cumulentgenfncross}  
\end{equation}
where $\left< r^j \right>_c$ is the $j$th cumulant.  From equation
(\ref{distpart10}) it can be seen that $F_n(p)$ is a sum of
terms containing ${n \choose m } m^s p^m$ where $s$ is some
positive integer.  These can be simply evaluated to give an
expression for $F_n(p)$ and hence $\rho_p$.  As for the
occupation partial survival, $F_n(p) = \rho(p)^n=\exp[-\theta(p)n]$ for all
$n$ even though this is not true for $P_{n,m}$.

\begin{figure} 
\narrowtext \centerline{\epsfxsize\columnwidth
\epsfbox{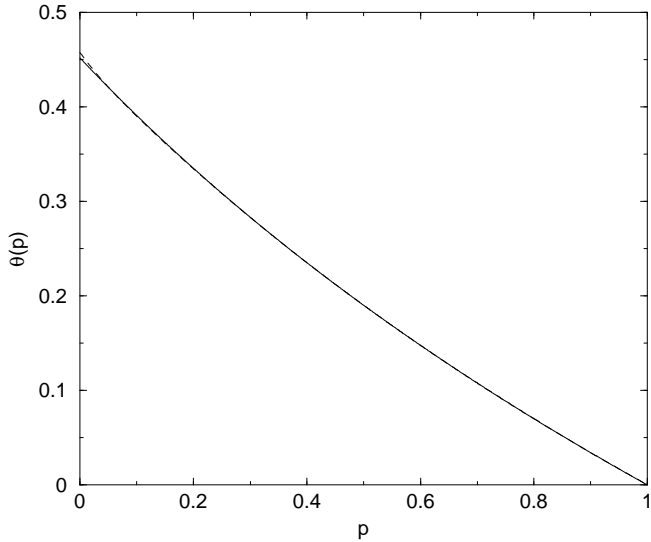}}
\caption{Plot of $\theta(p)$ against $p$ for the process
$\psi_i=(\phi_i+\phi_{i-1})/\sqrt{2}$.  The solid line is the 
exact result and the dashed line is the result of the correlator
expansion.  They differ by a maximum of $0.00576$ (at
$p=0$).  }
\label{fcrossings_satyassimpleprocess}

\end{figure}

For rough processes, the continuum partial survival is the
same as persistence, since any crossing entails an infinite
number of crossings and thus non-survival.  For smooth
processes however, calculation of $\rho(p)$ and hence that of 
$\theta(p)=-ln[\rho(p)]$ is non-trivial.
For the random acceleration problem the exact result is
\cite{burkhardtsnicepaper},

\begin{equation}  
\theta(p) = {1 \over 4} \left( 1-{6 \over \pi} \sin^{-1} {p \over 2}  \right).
\label{dummy26}  
\end{equation}
Whilst for the intrinsically discrete process (equation
(\ref{satyasprocess})) it is
\cite{majumdarpsieqphiiplusphiiminus1},

\begin{equation}  
\theta(p) = \ln \left({ \sin^{-1} \sqrt{1-p^2} \over \sqrt{1-p^2} }\right).
\label{dummy27}  
\end{equation}
Figure \ref{fcrossings_satyassimpleprocess} shows this
result and the raw series result for comparison.

\begin{figure} 
\narrowtext \centerline{\epsfxsize\columnwidth
\epsfbox{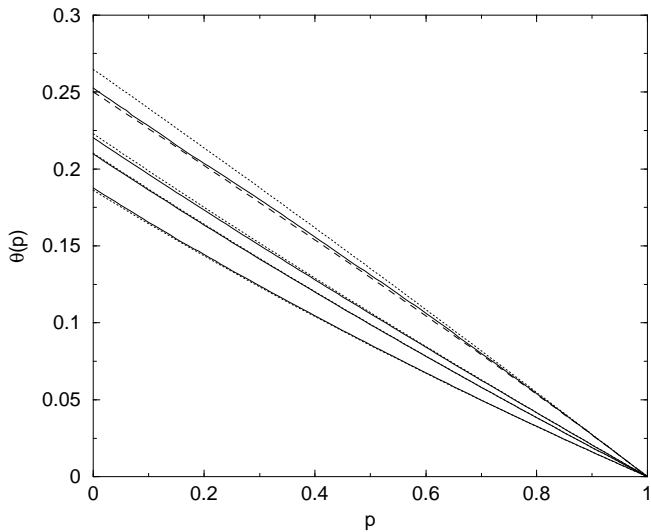}}
\caption{Plot of $\theta(p)$ against $p$ for (top to bottom) 
the $d^n x/dt^n = \eta(t)$ process with $n=2$ (random acceleration), 
$n=3$, $n=4$, and $n \to \infty$ (equivalent to diffusion in 2 dimensions).  
For $n=2$ (top curves) the exact result (\ref{dummy26}) 
is shown (dashed line) along with
the IIA (dotted) and the Pad\'e with 1 constraint (solid
line).  For the other cases, the IIA (dotted) and Pad\'e
with 2 constraints (solid line) are shown.  The correlator
results are an average of suitable Pad\'es of order 10 to 7.
Note that the IIA is rather inaccurate for the random
acceleration, but improves as $n$ increases. }
\label{fcrossings_ddxanddddx_ctexpandIIA}

\end{figure}

\begin{figure} 
\narrowtext \centerline{\epsfxsize\columnwidth
\epsfbox{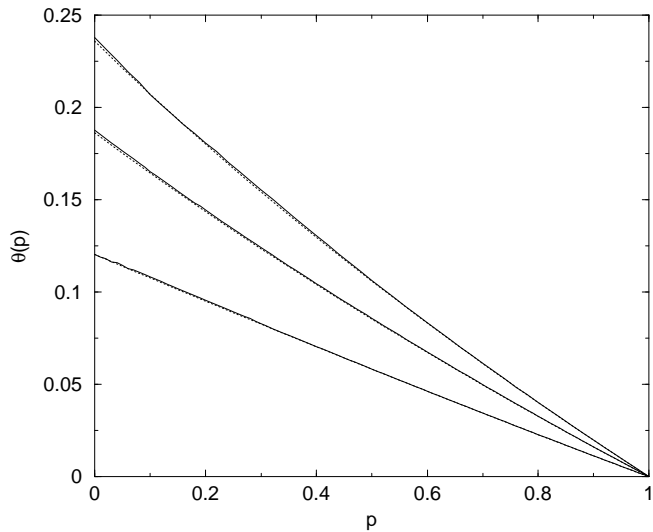 }}
\caption{Plot of $\theta(p)$ against $p$ for diffusion in 1,
2 and 3 dimensions (bottom to top).  The IIA results
(dotted) and the correlator results (solid lines) are
plotted.  The correlator results are an average of suitable
Pad\'es of order 10 to 7 for 1 and 2 dimensions and 6 to 5
for 3 dimensions.  }
\label{fcrossings_diffusion123_ctexpandIIA}

\end{figure}

For general processes we apply the constraint to the series that
$\rho(p)|_{\Delta\!T=0} =1$.  For sufficiently smooth processes,
as before, the first correction to $\theta(p)$ near
$\Delta\!T =0$ will be of order $\Delta\!T^2$.  We 
apply this additional constraint to the appropriate processes and 
the corresponding continuum results are shown in figures
\ref{fcrossings_ddxanddddx_ctexpandIIA},
\ref{fcrossings_diffusion123_ctexpandIIA}.

  \widetext

  \widetext

  \widetext

The results from the matrix method partial survival for the
random walk and random acceleration agree with those of the
correlator expansion term by term to within the numerical
error of the matrix method.  Also, using $F_n(p)$ as a
generating function, we find that the first two cumulants
agree term by term to 10th order with the results of section
\ref{The Variance of the Number of Crossings}.
We are also able to calculate higher cumulants.  We have
also used the method of steepest descents to calculate
$\rho(r)$ from $\rho(p)$ as a further cross-check.

This evaluation of the crossing partial survival completes
our calculations.

\section{Conclusion} \label{Conclusion}

In this paper we have used the correlator expansion to
calculate several properties of an arbitrary discrete or
discretely sampled Gaussian Stationary Process.  The
expansion in powers of the correlator works well when the
variables are relatively weakly correlated.  For stronger
correlations the series expansion does not converge.  We are
however able, for the case of an underlying continuous and
sufficiently smooth process, to extrapolate our results all
the way to the continuum by using the Pad\'e Approximant
with two constraints.  Thus even in the continuum we have
been able to calculate the persistence exponents, the
occupation-time exponents and the partial survival of
crossings exponents to high precision. These results compare well 
with those of the Independent Interval Approximation, the other 
general method.  In most cases they are more accurate, however, 
and they also give an estimate of the error of the result which the IIA
does not.  Furthermore, by calculating higher orders the
results may be improved. We believe we have demonstrated convincingly 
that the correlator expansion is the method of choice for calculating 
persistence properties of Gaussian stationary processes. 

\section{Acknowledgements}

GE thanks Andrew Stephenson for useful discussions, and EPSRC(UK)  
for a Research Studentship.

\end{document}